\documentclass{article}
\usepackage[frozencache,cachedir=.]{minted}
\usepackage{hyperref}
\usepackage{tikz}
\usepackage{booktabs}
\usepackage{csquotes}
\usepackage{colortbl}
\usepackage{listings}
\usepackage{authblk}
\usepackage[font=bf]{caption}
\newcommand*\circled[1]{\tikz[baseline=(char.base)]{
            \node[shape=circle,draw,inner sep=2pt] (char) {#1};}}
\usepackage[a4paper, margin=2cm]{geometry}

\title{Towards a high-performance AI compiler with upstream MLIR}
\author{\small Renato Golin}
\author{\small Lorenzo Chelini}
\author{\small Adam Siemieniuk}
\author{\small Kavitha Madhu}
\author{\small Niranjan Hasabnis}
\author{\small Hans Pabst}
\author{\small Evangelos Georganas}
\author{\small Alexander Heinecke}

\affil{\small Intel Corporation}
\date{}

\begin{document}

\maketitle

\begin{abstract}
This work proposes a compilation flow using open-source compiler passes to build a framework to achieve \textit{ninja} performance from a generic linear algebra high-level abstraction. We demonstrate this flow with a proof-of-concept MLIR project that uses input IR in Linalg-on-Tensor from TensorFlow and PyTorch, performs cache-level optimizations and lowering to micro-kernels for efficient vectorization, achieving over 90\% of the performance of ninja-written equivalent programs. The contributions of this work include: (1) Packing primitives on the tensor dialect and passes for cache-aware distribution of tensors (single and multi-core) and type-aware instructions (VNNI, BFDOT, BFMMLA), including propagation of shapes across the entire function; (2) A linear algebra pipeline, including tile, fuse and bufferization strategies to get model-level IR into hardware friendly \textit{tile} calls; (3) A mechanism for micro-kernel lowering to an open source library that supports various CPUs.
\end{abstract}

%\linenumbers
\section{Introduction}
Production high-performance code often needs to use highly-optimized libraries to achieve acceptable performance on modern hardware. Using generic micro-kernel libraries allows users to combine low-level calls into larger operations without worrying about the \textit{``last mile''} optimizations or optimal hardware utilization. However, designing highly-optimized libraries requires intricate knowledge of the problem space and the target architecture, thus leading to lack of generality. Furthermore, mapping those kernels on the existing applications is still a sizeable challenge, inaccessible to most application programmers~\cite{10.1145/3372266}.

Other AI compiler frameworks (ex. IREE~\footnote{https://github.com/openxla/iree}) have presented opportunities to accelerate application code by combining language extensions, graph sharding and data reordering, IR compiler memory bandwidth and compute density optimizations, and super-optimized libraries. However, due to the complexity of those frameworks and the vast domains they need to cover, support for efficient execution gets limited by common high-level patterns, leading to an explosion of problem-specific kernel implementations.

In our previous work on implementing a micro-kernel library for the Tensor Processing Primitives (TPP)~\cite{10.1145/3458817.3476206}, we demonstrated that one can build a comprehensive set of deep learning and high-performance algorithms and achieve state-of-the-art performance by selecting appropriate micro-kernels for the suitable tensor shapes. We have implemented those primitives in the open-source micro-kernel library named libxsmm~\footnote{https://github.com/libxsmm/libxsmm}, which can reach over 90\% performance of the achievable peak (hand-written assembly).

This work\footnote{https://github.com/plaidml/tpp-mlir} builds on the success of TPP's state-of-the-art performance of various algorithms on a variety of CPUs by bringing a set of high-level linear algebra compiler passes to automatically choose the correct TPP operations, in the right order, with the suitable flags, including packing tensors and adjusting iteration spaces for optimal traversal and full utilization of hardware resources. More importantly, it is possible (and desirable) to achieve this goal by having those passes in LLVM upstream, while adding a small low-level layer (dialect and conversion passes) specific to the library being used.

Our compiler is based on the well-known MLIR~\cite{10.1109/CGO51591.2021.9370308} technology by extracting programs from existing high-level frameworks (such as TensorFlow and PyTorch) through our tensor manipulation passes, into low level library and hardware dialects for further lowering.
This work exposes compiler heuristics via command line flags, allowing users to identify what constraints work best for each case, and investigate the boundaries with which to create a cost model that would drive this automatically. The construction and utilization of this cost model is a subject for future work.

Our main contribution is to enable the ease of using frameworks and automatic compilers on high-level programs with the performance of \textit{``ninja written''} low-level libraries, taking advantage of the last drop of hardware performance on a selection of architectures without resorting to user-specified schedules, pragmas, hints or hand-crafted intrinsics and inline assembly.

\section{MLIR and the Linalg Dialect}

The MLIR compiler infrastructure is a project under the LLVM umbrella well-suited for multi-level IR rewriting. MLIR provides an intermediate representation (IR) with only a few concepts being
built, leaving most IR customizable. Such IR allows compiler developers to match the right abstraction level for their problems by introducing custom types, operations, and attributes. We use the static single-assignment (SSA) form, which is suitable for imperative programming styles that machine learning (ML) and high-performance computing (HPC) applications lower to.

A dialect is a basic structure that enables the MLIR to implement a stack of reusable abstractions, composed of operations, types, attributes, etc. Each abstraction encodes and preserves transformation validity preconditions directly in its IR, reducing the complexity and the cost of analysis passes.

Each dialect models a specific domain. For example, the Linalg dialect~\footnote{https://mlir.llvm.org/docs/Dialects/Linalg} captures linear-algebra operations on either tensor or buffer operands. Listing~\ref{lst:linalg_generic} shows a single layer of a multi-layer perceptron ML model, with a matrix multiplication followed by a \textit{``bias''} addition and rectifier (ReLU) implemented as \texttt{max(x, 0.0)}. The operations' semantics describe the computation in the \textit{``inner loop''} and define iteration order via \textit{``indexing\_maps''} and \textit{``iterator\_types''}.

As is common in the MLIR ecosystem, we originally developed our own (TPP) tile dialect, between Linalg on Tensors and our XSMM dialect, which allowed us to manipulate tiling, fusing, and bufferization on our terms. However, through upstream discussions on a tile dialect design, we have concluded that the upstream Linalg dialect should have such abstractions. We then updated the Linalg dialect with our operations and now use only Linalg for both whole-tensor and tile semantics, on tensors and memrefs. This allows us to further our mission to impact and reuse upstream high-level transformations for a common compilation infrastructure.

\begin{listing}[p]
\begin{center}
\begin{minipage}[]{\textwidth}
\begin{minted}[fontsize=\small, escapeinside=@@, mathescape=true]{cpp}
    // Affine maps M,K * K,N -> M,N
    #map-mk = affine_map<(d0, d1, d2) -> (d0, d2)>
    #map-kn = affine_map<(d0, d1, d2) -> (d2, d1)>
    #map-mn = affine_map<(d0, d1, d2) -> (d0, d1)>

    // A perfectly nested affine fused mltiply and accumulate operation (matmul)
    %0 = linalg.generic {
        indexing_maps = [@#@map-mk, @#@map-kn, @#@map-mn],
        // Reduction iterator type is the third, ie. ``d2'', which is the ``K'' dimension
        iterator_types = ["parallel", "parallel", "reduction"]
    }
    // Inputs are A and B matrices, C is the initialized of the output (generally zero).
    ins(%A, %B : tensor<128x256xf32>, tensor<256x512xf32>)
    // Output is the C matrix, here representing initialization (C+= A * B), where C can be zero
    outs(%C : tensor<128x512xf32>) {
            ^bb0(%in: f32, %in_1: f32, %out: f32):
              %3 = arith.mulf %in, %in_1 : f32
              %4 = arith.addf %out, %3 : f32
              linalg.yield %4 : f32
    } -> tensor<128x512xf32>

    // Affine maps element-wise & broadcast
    #map-ew = affine_map<(d0, d1) -> (d0, d1)>
    #map-bc = affine_map<(d0, d1) -> (d1)>

    // A binary operation on the output of the matmul above (ex. Bias Add)
    %1 = linalg.generic {
        indexing_maps = [@#@map-ew, @#@map-bc],
        iterator_types = ["parallel", "parallel"]
    }
    // Inputs are C and Bias matrices.
    // Note: the bias is a 1D vector being broadcasted to add element-wise.
    // Note: the C matrix is the initializer of the output, so it's in `outs`.
    ins(%BIAS : tensor<512xf32>)
    outs(%0 : tensor<128x512xf32>) {
            ^bb0(%in: f32, %out: f32):
              %4 = arith.addf %in, %out : f32
              linalg.yield %4 : f32
    } -> tensor<128x512xf32>

    // A unnary operation on the output of the binary above (ex. ReLU)
    %ZERO = arith.constant 0.000000e+00 : f32
    %2 = linalg.generic {
        // Element-wise parallel operation only uses MN maps
        indexing_maps = [@#@map-ew],
        iterator_types = ["parallel", "parallel"]
    }
    // Input is just the result above.
    // Note: the result is the initializer of the output, so it's in `outs`.
    outs(%1 : tensor<128x512xf32>) {
            ^bb0(%out: f32):
              %4 = arith.maximumf %out, %ZERO : f32
              linalg.yield %4 : f32
    } -> tensor<128x512xf32>

    return %2
\end{minted}
\end{minipage}
\caption{A simple multi-layer perceptron (MLP) layer represented in Linalg generic operations. The affine maps in ``indexing\_map'' describe the iteration space of each input, the ``iterator\_types'' the type of loop (parallel or reduction), while the inner region specifies the computation. The arguments (``ins'' and ``outs'') are tensors created beforehand.}
\label{lst:linalg_generic}
\end{center}
\end{listing}

\section{Compilation Strategy}

\begin{figure}[h]
    \centering
    \includegraphics[width=1\linewidth]{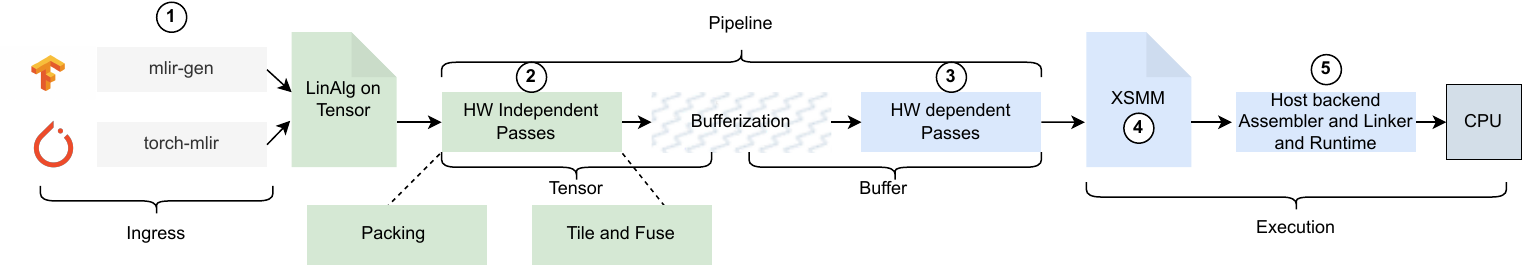}
    \caption{A simplified view of the proposed compiler strategy. In gray are external components, in green are the upstream compiler technology while in blue are the potentially downstream parts. Boundaries depend on which ingress format and which hardware abstractions are used. XSMM is our choice of CPU library, and OpenCL is a potential choice for GPU libraries. The proposal is equally valid with dialects and further compilers (ex. LLVM) down the line.}
    \label{fig:overview}
\end{figure}

Our compilation strategy in figure~\ref{fig:overview} is based on five main components: \circled{1} An ingress layer~\footnote{https://github.com/plaidml/mlir-generator}, that extracts MLIR from existing frameworks into Linalg-on-Tensor Intermediate Representation (IR); \circled{2} A high-level hardware-agnostic Linalg pipeline; \circled{3} A low-level lowering dialect; \circled{4} A pipeline for libxsmm; \circled{5} An execution strategy for the generated code, including runtime libraries and wrappers.

The ingress layer is based on external frameworks: IREE for TensorFlow models and torch-mlir~\footnote{https://github.com/llvm-project/torch-mlir} for PyTorch Dynamo. On top of that, we have created a tool named \textit{mlir-gen} that replicates the IR that IREE generates from TensorFlow but is extensible enough to generate various shapes, number of layers, size of matrices, type packing, etc. This IR generator aims at providing dynamic inputs to our compiler tests and replaces our dependency on IREE, as seen in figure~\ref{fig:overview}.

The second component is a Linalg-based pipeline built by composing upstream building blocks. Following previous work~\cite{vasilache2022composable}, we have implemented a focused pipeline, using existing passes with our new operations (\texttt{tensor.pack} and \texttt{tensor.unpack}) and their respective transformations.

%Our implementation could be improved when considering a fully flexible Linalg optimizing pipeline. Still, we show that it is possible to make such a pipeline only using Linalg passes and reasonable heuristics to achieve high performance by combining compiler passes with minimal vendor-optimized routines. The idea is to leverage compilers for what they are good at: fusion, tiling, and data layout while relying on vendor-optimized routines for optimizations like vectorization and instruction selection.
While our implementation is not yet a fully flexible Linalg optimizing pipeline, we demonstrate the viability of such a pure Linalg-based approach. In essence that means, relying on only using Linalg passes and reasonable heuristics to achieve high performance by combining compiler passes with minimal vendor-optimized routines. The idea is to leverage compilers for what they are good at: fusion, tiling, and data layout while relying on vendor-optimized routines for optimizations like vectorization and instruction selection.

Our core mission is to have a Linalg-based optimization pipeline upstream in MLIR. Our first step in this direction was the upstreaming of operations (\texttt{tensor.pack}, \texttt{tensor.unpack} and several Linalg named operations) and creation and improvement of upstream passes (packing, tiling, fusion).

The fourth component is an in-house dialect (XSMM) that plays the role of interfacing with our ``last-mile'' library: \textit{libxsmm}. The XSMM dialect does not need to be upstream, as it represents a third-party library and its semantics. As such, we show that it's feasible to have an upstream Linalg-based compiler for the high-level transformations and a downstream target-specific dialect to do low-level transformations. XSMM dialect is used to perform transformations that are specific to the library, not necessarily the final target architecture. For example, it exposes fused library calls that can implement a whole MLP layer (GEMM + Binary + Unary), saving on loads/stores, register renaming, cache flushes, and buffer allocation. The high-level Linalg IR does not need to know or care about this.
Lowering Linalg on memrefs into other hardware specific dialects (such as \texttt{vector} or GPU target dialects) is subject for future work.

The fifth and last component is a simple runtime wrapper for libxsmm and a just-in-time compiler infrastructure to get executable code. We reuse the upstream \texttt{mlir-cpu-runner}, extending it to include our dialects and passes as well as to generate a benchmark loop for the kernels being tested (using a local \texttt{perf} dialect that we're discussing how to upstream).

\subsection{Why Linalg on Tensors?}

Different frameworks (e.g. TensorFlow, PyTorch) implement their own MLIR dialects (StableHLO, Torch) with semantics equivalent to their internal graph formats. There are other high-level dialects (ex. TOSA\footnote{https://mlir.llvm.org/docs/Dialects/TOSA}) but they also have their individual semantics, which makes it hard to work on a common optimization infrastructure.

To avoid this complexity, compilers aim at converting those dialects into a common \textit{``compiler IR''}: Linalg on Tensor. These two dialects can describe a substantial part of ML applications with a small number of generic operations (as seen in Listing~\ref{lst:linalg_generic}) and are very amenable to transformations.

Being at a tensor \textit{value} semantics, where each new operation materializes a new tensor, allows us to perform most of the optimization we need, namely packing, kernel fusion, and tiling, without having to worry about memory constraints. In addition, there are upstream passes that we want to reuse that run on either Linalg or its interfaces, many of those to which we contributed substantially.

This set of dialects, with tiled and fused operations is then bufferized by the \textit{one-shot} bufferization pass, cleaned, canonicalized and further lowered to low level dialects such as XSMM, where library/hardware specific passes can operate on an already memory-friendly shape.

\section{Compiler Passes}

The compiler pipeline consists of two parts: \circled{2} high-level optimizations, generic to all targets and based on cost models, heuristics-based decisions, and graph-rewrites, and \circled{3} low-level optimizations, specific to each chosen target, primarily adjusting the high-level decisions to the target expectations (low-level dialects, function calls, intrinsic, etc).

High-level passes like packing, tiling, and fusing are hardware-agnostic (with hardware-specific costs) and aim to reduce round-trip to memory by providing aggressive fusion strategies around contraction-like operations (i.e., fuse element-wise operations with matmul). Tiling exposes scalar or parallel loops around operations that can map 1:1 with micro-kernel operations in libxsmm (exposed by the XSMM dialect), but also fits common CPU and GPU code generation patterns (using \texttt{vector}, \texttt{gpu} and other close-to-hardware dialects). 

Low-level passes can then be done on the specific low-level dialects (such as XSMM and vector), where target-specific transformations can be done at a representation closer to the hardware/library.

\subsection{Pack, Unpack, and Propagation}

Packing (or data-blocking) is a well-known transformation in high-performance libraries that copies a non-contiguous block of data to a contiguous block in memory to reduce the number of TLB entries required to access each page. When copying data, packing rearranges block elements to decrease the stride between consecutive accesses, improving spatial locality and cache behavior.

\begin{figure} [ht]
    \begin{minipage}{\textwidth}
    \includegraphics[width=1\linewidth]{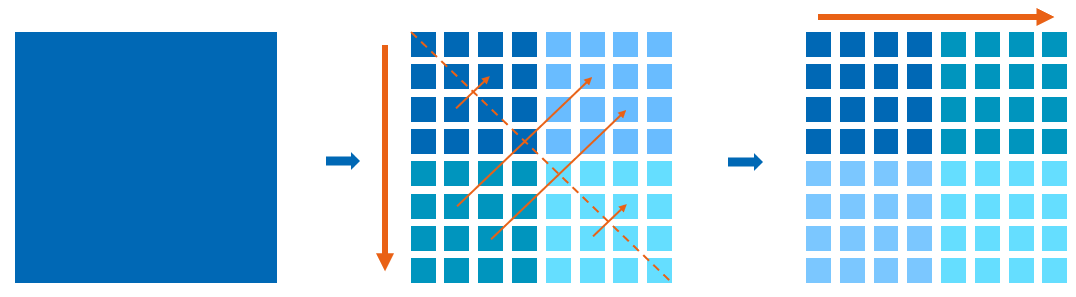}
    \end{minipage}
    \break
    \begin{minipage}{\textwidth}
    \includegraphics[width=1\linewidth]{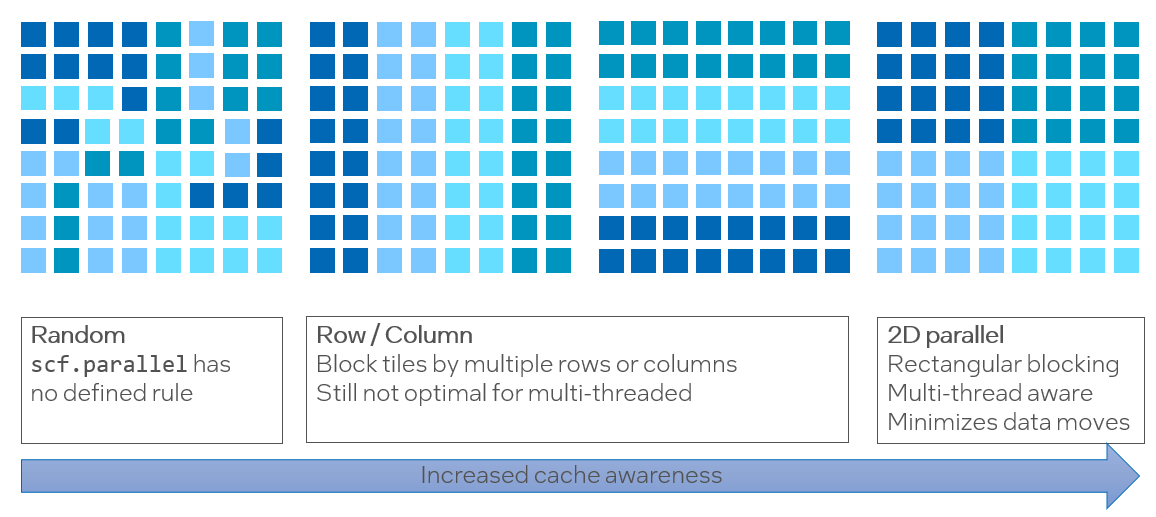}
    \end{minipage}
    \caption{Packed layout for GEMM operation. After tiling (smaller square), the tiles are transposed whole (\textit{``block-transpose''}). For optimal multi-threaded locality we also group different blocks (single-colored areas) for each thread.}
    \label{figure:packing-shapes}
\end{figure}

Bringing it into the compiler has advantages, as we can propagate the layout through the IR instead of paying the price for every invocation. We perform packing by introducing two new operations in the tensor dialect: \texttt{tensor.pack} and \texttt{tensor.unpack}. The former takes a tensor as input and produces a re-layout tensor as output, while the latter undoes the packing, bringing the tensor back to the original layout. Table~\ref{figure:packing-shapes} shows one of the layouts we use for different matmul operations. 

Currently, we employ very simple heuristics around packing: select an optimal tile size and divide the outer dimensions around it, reordering the blocks of the second tensor to improve cache locality. The optimal tile size is currently fixed but with the option to allow users to override it. A cost-model-based heuristics to select the best tile size for a given architecture is subject to further work.

Each operation is packed as long as its iteration space has more iteration than the packing factor encoded in the pass. Usually, if the iterations on each dimension are less than a certain architecture-specific amount, the computation fits in the cache, and packing is unnecessary.

\begin{figure} [ht]
    \begin{minipage}{\textwidth}
    \includegraphics[width=1\linewidth]{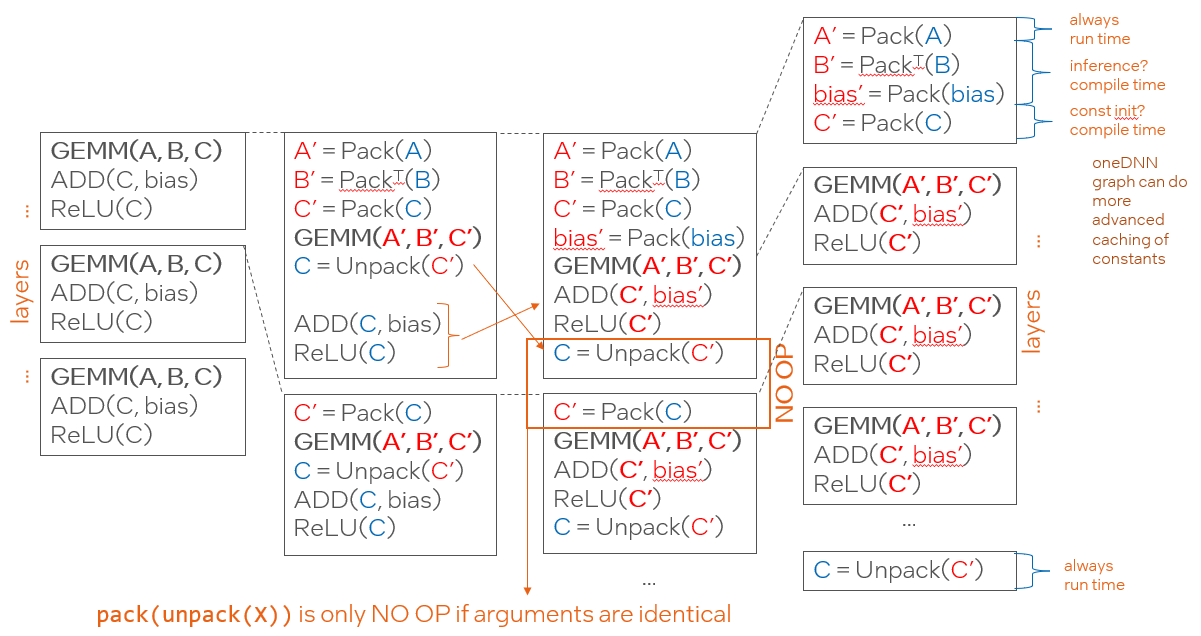}
    \end{minipage}
    \caption{Pack propagation through a multi-layer model. Packed GEMMs propagate their layout to the following element-wise operations, exposing canonicalization in between layers to elide all intermediate packs and unpacks, leaving only the initial packs and final unpack.}
    \label{figure:packing-propagation}
\end{figure}

To amortize the cost of packing, packing operations are propagated through element-wise operations, and a pair of \texttt{unpack(pack(t))} are folded away when possible (figure~\ref{figure:packing-propagation}). This allows us to be naive on the initial packing, just looking at large matrix multiply and contractions, then expand the same shapes throughout the following operations reaching the next stages (following layer, next step in the same layer), and remove all intermediate packing operations except the first ones on inputs and last ones on return values.

Packing constants (weights and biases on inference models) can be done at compile time if we have access to the data in the model (for example as a constant global or an \texttt{arith.constant}). However, more advanced folding~\cite{li2023onednn} can be done at run time even on models where the constant data is passed as arguments by packing and computing the first time, caching and reusing the subsequent times.
Moreover, zero-initialized buffers (for example, the output C matrix tile for matmul) can be just reshaped at compile time, since it's only a \textit{splat} of zeroes onto an arbitrary shape.

\subsection{Tile and Fuse}

We use upstream building blocks to assemble our tiling and fuse pass. The main idea is to consider contraction operations and fuse them along the parallel dimensions with consumers' or producers' element-wise operations.

As in Listing~\ref{lst:tile_and_fuse}, we fuse along the \texttt{M} and \texttt{N} parallel loop. Using a bottom-up traversal, we consider each contraction and collect possible fusable consumers or producers looking at their iteration domain. This effectively creates a cluster of element-wise operations around a contraction.
We require each operation to belong to a single cluster domain, and we prevent recomputation by avoiding duplicating operations that belong to multiple fusion domains. 

\begin{listing}[hb]
\begin{minipage}[]{\textwidth}
\begin{minted}[fontsize=\small, escapeinside=@@, mathescape=true]{cpp}
// Convert tile-wise operations
for(MB, NB) {
  for (KB) {
    C[MB][NB][mb][nb] += A[MB][KB][mb][kb] * B[NB][KB][kb][nb]
  }
  C[MB][NB][mb][nb] = add(C[MB][NB][mb][nb], bias[mb][nb])
  C[MB][NB][mb][nb] = max(C[MB][NB][mb][nb], 0)
}

// Into a parallel BRGEMM "tile" op + element-wise tail ops
parallel(MB, NB) {
  -> extract { A, B, C } x [MB][NB][KB] as apporpriate
  // Note: this is a batch-reduce GEMM into a "tile"
  C[mb][nb] += A[KB][mb][kb] * B[KB][kb][nb]
  C[mb][nb] = add(C[mb][nb], bias[mb][nb])
  C[mb][nb] = max(C[mb][nb], 0)
  <- insert into C[MB][NB]
}
\end{minted}
\end{minipage}
\caption{The tile and fuse pass will materialize the two parallel loops \texttt{MB} and \texttt{NB}, because both are parallel outer dimensions, i.e., there are no loop carried dependencies between the tile operations.}
\label{lst:tile_and_fuse}
\end{listing}

\subsection{Lowering to Hardware Dialects}

After the hardware independent passes, we run the upstream bufferization pass to move from a value-based \texttt{tensor}s to memory buffer \texttt{memref}s. After bufferization, the Linalg on buffers IR, now operating in tile shapes inside parallel loops, is lowered to our XSMM dialect, which acts as an interface to libxsmm. After low-level dialect-specific optimization passes, XSMM is lowered to function calls that invoke libxsmm through a simple C runtime.

\subsubsection{The XSMM dialect}

The XSMM dialect maps the behavior of the libxsmm library and enables library-specific optimizations, for example, call fusion.

The libxsmm library is a JIT-ing library and is split into two stages: \textit{dispatch} and \textit{invoke}.
The dispatch stage receives the shapes of the buffer, leading dimensions, broadcast, and fusion flags and compiles the micro-kernel in memory, returning a pointer into its implementation. The second time a \textit{dispatch} function is called, it just returns a cached pointer to the same implementation.
The invoke stage calls that function pointer with the actual tensor data (usually a tile into a larger buffer with the appropriate strides), which computes the operation, writing the result to the output buffer.

XSMM exposes only five operations: \texttt{unary}, \texttt{binary}, \texttt{gemm}, \texttt{brgemm} and \texttt{fused\_brgemm}. \texttt{unary} are element-wise unary operations like ReLU, but also broadcasts, transposes and reductions. \texttt{binary} are element-wise binary operations like add or multiply. \texttt{gemm} represents General Matrix Multiplications mirroring the BLAS interface; \texttt{brgemm} is a more powerful abstraction that carries an extra reduction dimension on the input operands, allowing reducing tiles of A and B in the same C tile. Finally, \texttt{fused\_brgemm} allows register fusion between BRGEMM and an element-wise prologue and epilogue.

The last operation above demonstrates the power of library-specific (not just hardware-specific) compiler optimizations. To allow for better low-level optimization, libxsmm implements a fused BRGEMM, which can add the following arbitrary binary and unary operations to the BRGEMM, including broadcast semantics.

\begin{listing}[h]
\begin{minipage}[]{\textwidth}
\begin{minted}[fontsize=\small, escapeinside=@@, mathescape=true]{cpp}
// Convert multiple XSMM calls
%3 = xsmm.unary.dispatch zero [...] flags = (none)
%4 = xsmm.brgemm.dispatch [none] flags = (none)
%5 = xsmm.binary.dispatch add [...] flags = (bcast_col_in0)
%6 = xsmm.unary.dispatch relu [...] flags = (none)
scf.parallel (MB, NB) {
    %subview_A = memref.subview ... // into A
    %subview_B = memref.subview ... // into B
    %subview_C = memref.subview ... // into C

    // C[MB][NB] = { 0.0 }
    xsmm.zero(..., %3, %subview_C)
    // C[MB][NB] = BRGEMM(A[MB][NB], B[MB][NB], C[MB][NB])
    xsmm.brgemm(data_type = f32, %4, %subview_A, %subview_B, %subview_C, %c0)
    // C[MB][NB] = ADD(broadcast(Bias[NB]), C[MB][NB])
    xsmm.binary add(..., %5, %BIAS, %subview_C, %subview_C)
    // C[MB][NB] = ReLU(C[MB][NB])
    xsmm.unary relu(..., %6, %subview_C, %subview_C)
}

// Into a single fused one with all flags
%3 = xsmm.fused_brgemm.dispatch [...][add,relu] flags = (beta_0)  
    binary_flags = (bcast_col_in0) unary_flags = (none)
scf.parallel(MB, NB) {
    %subview_A = memref.subview ... // into A
    %subview_B = memref.subview ... // into B
    %subview_C = memref.subview ... // into C

    // C[MB][NB] = { 0.0 }
    // C[MB][NB] = BRGEMM(A[MB][NB], B[MB][NB], C[MB][NB])
    // C[MB][NB] = ADD(broadcast(Bias[NB]), C[MB][NB])
    // C[MB][NB] = ReLU(C[MB][NB])
    xsmm.fused_brgemm(..., %3, %subview_A, %subview_B, %subview_C, %BIAS %c4)
}
\end{minted}
\end{minipage}
\caption{XSMM pass that fuses multiple XSMM calls into a fused call to improve register and memory usage.}
\label{lst:xsmm_fusion}
\end{listing}

\subsection{Parallelism}

We support parallelism using OpenMP by converting \texttt{scf.forall} into \texttt{scf.parallel} to expose CPU thread parallelism by running the upstream OpenMP conversion passes and linking with LLVM's OpenMP library.

We force the OpenMP affinity to guarantee the correct distribution of the tiles onto the appropriate threads with the setting \texttt{KMP\_AFFINITY} to \texttt{granularity=fine, verbose, compact,1,0}.

\subsubsection{2D parallelism}

We perform 2D parallelism transformation on the loops around the XSMM dialect (see figure~\ref{figure:packing-shapes}. This transformation distributes the OpenMP threads across the appropriate tile regions to increase reuse and minimize core-to-core traffic in a multi-core chip by best possible leveraging matrix multiplication's algorithmic properties. Different tile region shapes are optimal for different number of threads, micro-architecture extensions and XSMM-specific behaviour, so this pass is done at the low-level pipeline to extract the most of it. 

This pass further tiles the outer parallel dimension loops into smaller iteration spaces (multiple of the number of threads) and creates sequential loops inside each thread, guaranteeing the execution of all tiles in the same \textit{group} to be on the same thread.
Getting that balance right (memory bandwidth, caching, compute distribution) is not trivial.

Compiler command line options are introduced to select the \textit{best known} factor for each run on our benchmarks. However, as above, this technique is only being used to collect data on costs to support high-level heuristics decisions and will be subject to future work.

\subsubsection{AMX tile configuration hoisting}

The Intel AMX matrix unit needs to be configured via a status register before usage and reset after usage. This configuration is expensive as it resets the unit twice inside the inner loop.

Without further information, libxsmm must setup and reset around every loop according the ABI as both the configuration of the status register and the actual registers are defined as volatile. To avoid this extra cost, the compiler splits the \texttt{GEMM} operation into a triplet: \texttt{tile config} + \texttt{GEMM} + \texttt{tile reset}, and because the tile operations do not use any data from inside the inner loop (all dominators are outside), we can hoist the call outside of the loop.

\begin{figure} [ht]
    \begin{minipage}{\textwidth}
    \includegraphics[width=1\linewidth]{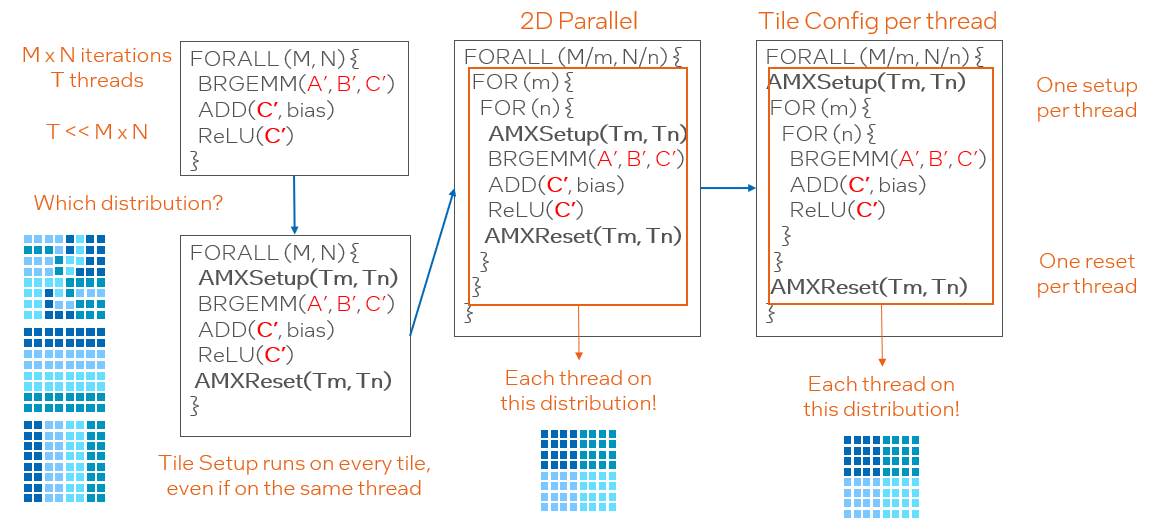}
    \end{minipage}
    \caption{Each layer is executed across all cores (data parallelism). For data locality, we block each thread within a single block (of tiles) within the original matrix (see figure~\ref{figure:packing-shapes}). To amortize the cost of using the matrix extension on Sapphire Rapids, we hoist the setup and reset calls within each thread.}
    \label{figure:2d-parallel-amx}
\end{figure}

However, as shown in figure~\ref{figure:2d-parallel-amx}, we must setup and reset the tiles at least once per thread, so we only hoist the tile calls past the sequential loops, not past the outer parallel loop (that will be executed on different threads).

\section{Results}

We compare our compiler's performance against libxsmm hand-written code (from its \texttt{libxsmm-dnn} repository), which, after micro-architecture analysis, is known to reach higher than 90\% of the achievable total performance of each CPU we target. Our compiler emits calls to the same micro-kernels, so the delta between the hand-written and compiler versions is due to the automatic parallelization of the input IR and not the low-level hardware optimizations. 

The benchmarks use an MLP model to demonstrate the critical optimization in machine learning, not as a measure of output performance in existing ML models (such as ResNet, DLRM, and BERT). Our test kernels mimic a 3-layer MLP with batch size 256 and hidden sizes 1024. We consider only inference where weights and biases are constants. Using such motifs for workload abstraction is common practice. It has been used by Google when introducing the TPU hardware or Meta when proposing the DLRM benchmark.

The CPU architectures we run on from AWS instances are:
\begin{itemize}
    \item (c6a): AMD(R) EPYC 7R13 (Zen3), supporting AVX2 (BF16 emulated)
    \item (c6i): Intel(R) Xeon(R) Platinum 8375C (CLX), supporting AVX512 (BF16 emulated)
    \item (c7i): Intel(R) Xeon(R) Platinum 8488C (SPR), supporting AMX (BF16 native)
    \item (c7a): AMD(R) EPYC 9R14 (Zen4), supporting AVX512\_bf16 (BF16 native)
    \item (c7g): AWS Graviton 3 (Gvt3), ARM V1 core supporting SVE (BF16 native)
\end{itemize}

The c6 architectures do not support native BF16 float types and were benchmarked to demonstrated our strategy works for 32-bit floating point on older hardware. As the BF16 types are emulated, we achieve lower performance than 32-bit, which is expected, since we cannot use 16-bit floats natively.

The c7 architectures all have native BF16 support and show our results in three different vendors. On Intel Sapphire Rapids (c7i), we use the AMX extension for 2D matrix operations; On AMD Genoa Zen4 (c7a) we use the AVX512\_bf16 extension; On AWS Graviton 3 (c7g), we use SVE's scalable vector BFMMLA extension for fast BF16 matrix operations.

We pin all our single-core benchmarks on core 3 and use OpenMP affinity from core 1 onward to avoid noise in benchmark results as core 0 is being used by the kernel.

\begin{figure} [hb]
    \begin{minipage}{0.19\textwidth}
    \includegraphics[width=1\linewidth]{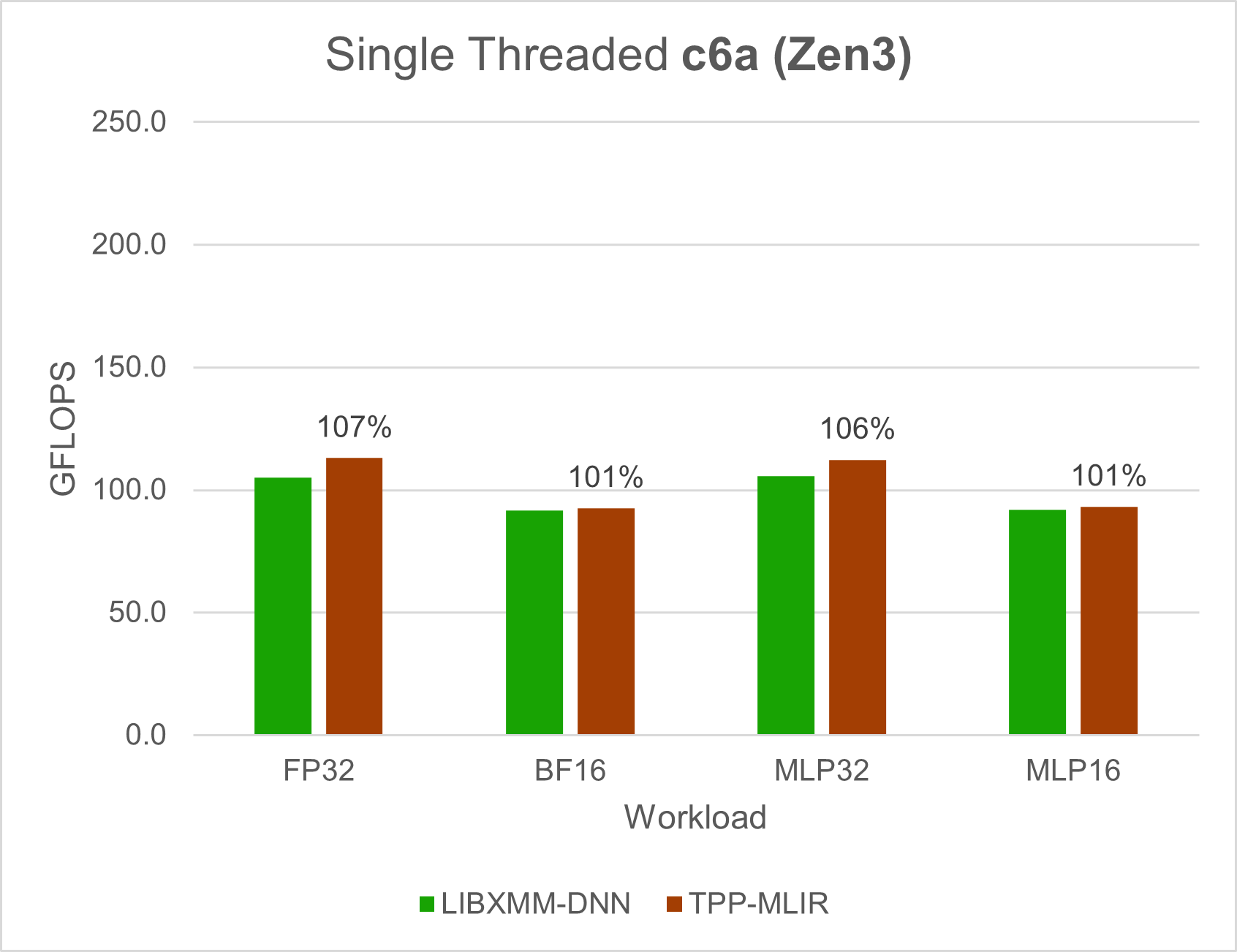}
    \end{minipage}
    \hspace{\fill} % note: no blank line here
    \begin{minipage}{0.19\textwidth}
    \includegraphics[width=1\linewidth]{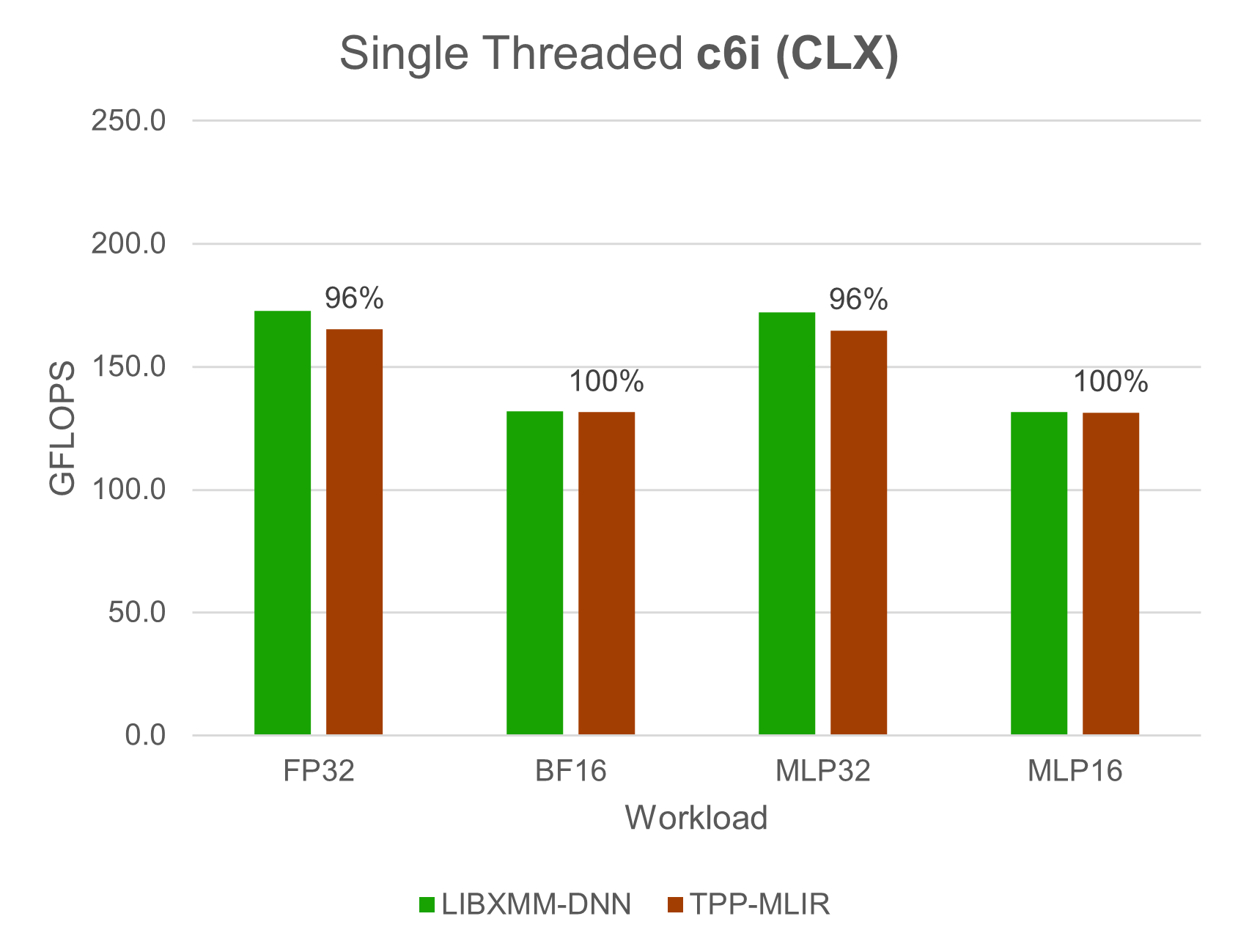}
    \end{minipage}
    \hspace{\fill} % note: no blank line here
    \begin{minipage}{0.19\textwidth}
    \includegraphics[width=1\linewidth]{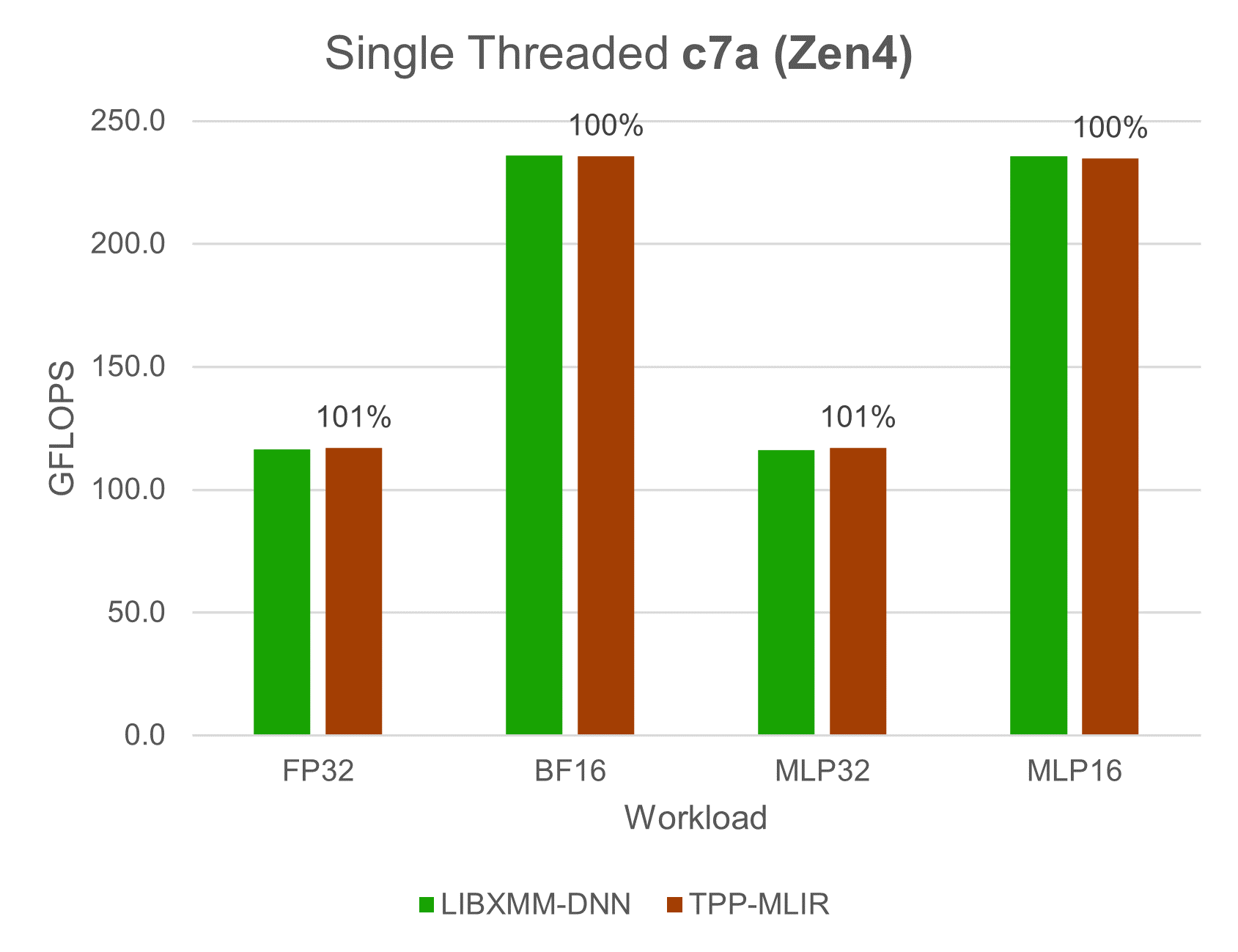}
    \end{minipage}
    \hspace{\fill} % note: no blank line here
    \begin{minipage}{0.19\textwidth}
    \includegraphics[width=1\linewidth]{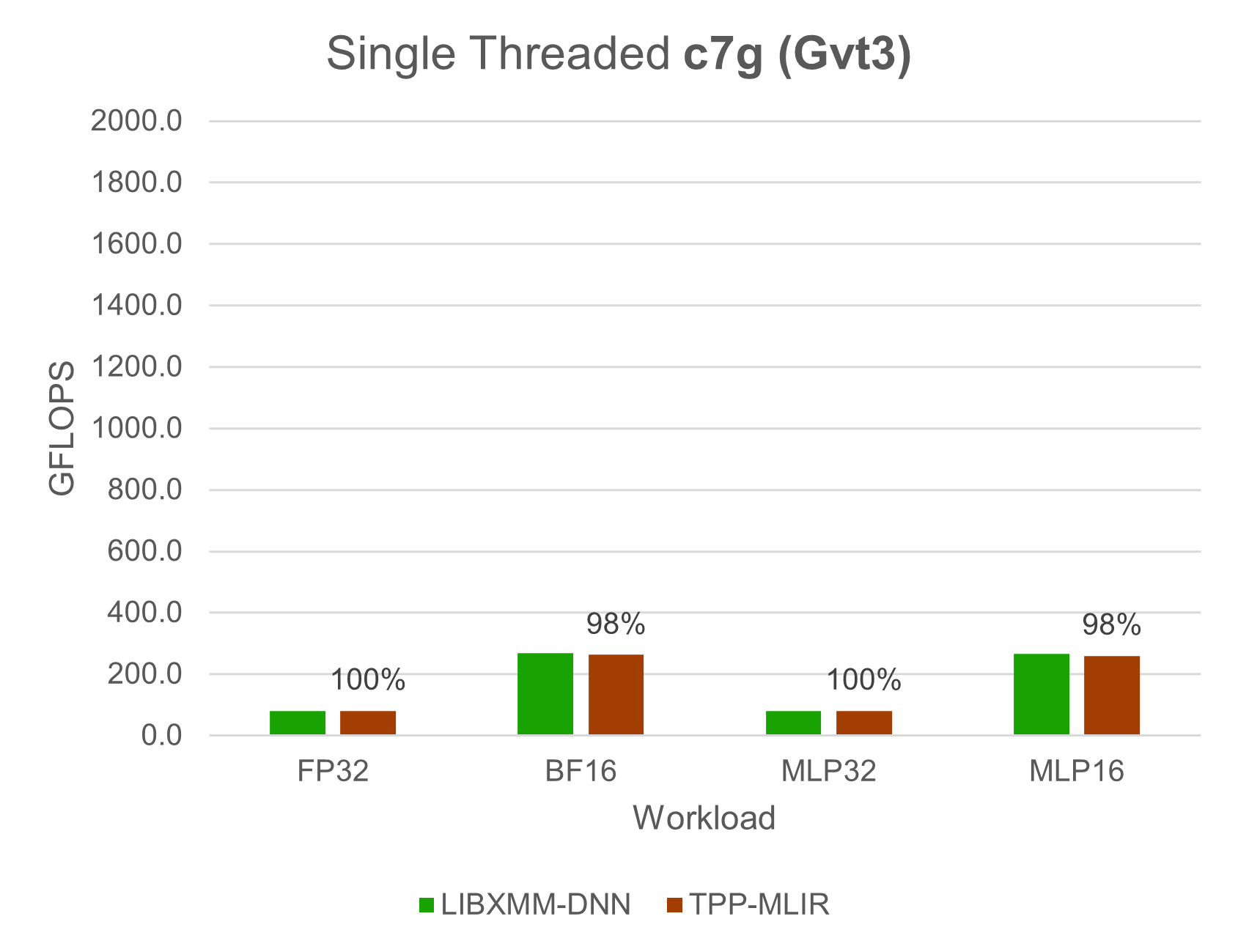}
    \end{minipage}
    \hspace{\fill} % note: no blank line here
    \begin{minipage}{0.19\textwidth}
    \includegraphics[width=1\linewidth]{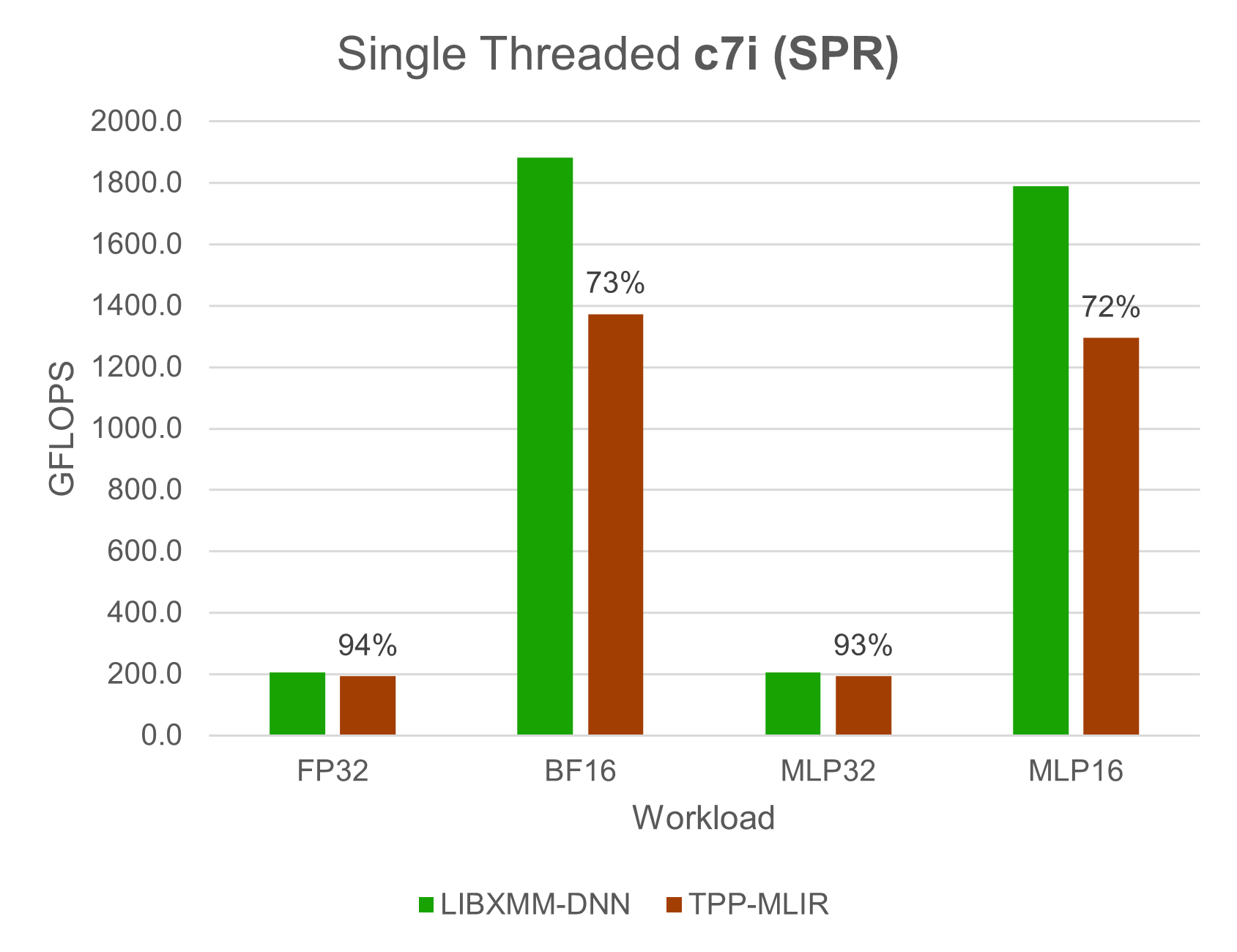}
    \end{minipage}

    \caption{Single-thread results for all CPUs. The compiler's performance is on par against all hand-written results except c7i (SPR), where compute density can be affecting memory bandwidth. Note, the scale on the first three plots are up to 250 GFLOPS, while the last two are up to 2 TFLOPS.}
    \label{fig:single-thread}
\end{figure}

\subsection{Single-thread performance}

Figure \ref{fig:single-thread} compares two types of runs:
\begin{enumerate}
    \item \textbf{LIBXSMM-DNN}, using hand-written C++ code calling libxsmm micro-kernels in the optimal shapes for maximum known performance. This is our baseline.
    \item \textbf{TPP-MLIR}, using our compiler to call the same micro-kernels above, but with the optimal configurations selected by the compiler. The model shapes here are in 4D, since the hand-written code does not do \textit{online} packing. This allows a fair performance comparison.
\end{enumerate}

On 32-bit floats (FP32), the compiler achieves the same overall performance as the hand-written code on all machines, with a deviation of less than 5\%. The compiler is slightly slower on Intel machines (c6i and c7i), equally fast on Arm (c7g) and slightly faster on AMD machines (c6a and c7a). This is due to the hand-written code being optimized on Intel machines and the compiler being a more generic approach.

This is not too surprising, given that small variations can be seen with different choices for loop order, tile sizes and tensor caching that \texttt{parlooper}~\cite{parlooper} finds with an exhaustive search. We have seen our compiler \textit{``get lucky''} a few times when comparing with hand-written code. Future work is needed to explore that space systematically and always produce the most efficient configuration for every target.

We have identified a further difference between the hand-written code and the compiler, where the former allocates a single (strongly aligned) \textit{arena} region to use as scratchpad, while the compiler relies on individual (not necessarily aligned) memory allocations. While this is unrelated to micro-kernel execution, it is a high-level decision that needs to be in the compiler's tool kit.
Further analysis and implementation of these fixes is subject to future work.

On 16-bit \textit{``brain''} floats (BF16), the picture requires a bit more explanation. Neither Zen3 nor CLX have native support for BF16, so we emulate it with older instruction sets, and therefore the performance is lower than FP32. However, they're still the same as the hand-written code on all machines except c7i. 

Sapphire Rapids (c7i) uses the AMX extension, which provides a much higher compute density than FP32, and therefore the effects of memory bandwidth are more apparent. Especially, if some of those loads and stores are done on unaligned memory, we can get a pathological case that is 4x slower than peak, which can easily lead to a 30\% performance degradation. Similar effects can be happening on Graviton 3's SVE extensions, but at a lower rate given the smaller compute density.

\begin{figure} [ht]
    \begin{minipage}{0.24\textwidth}
    \includegraphics[width=1\linewidth]{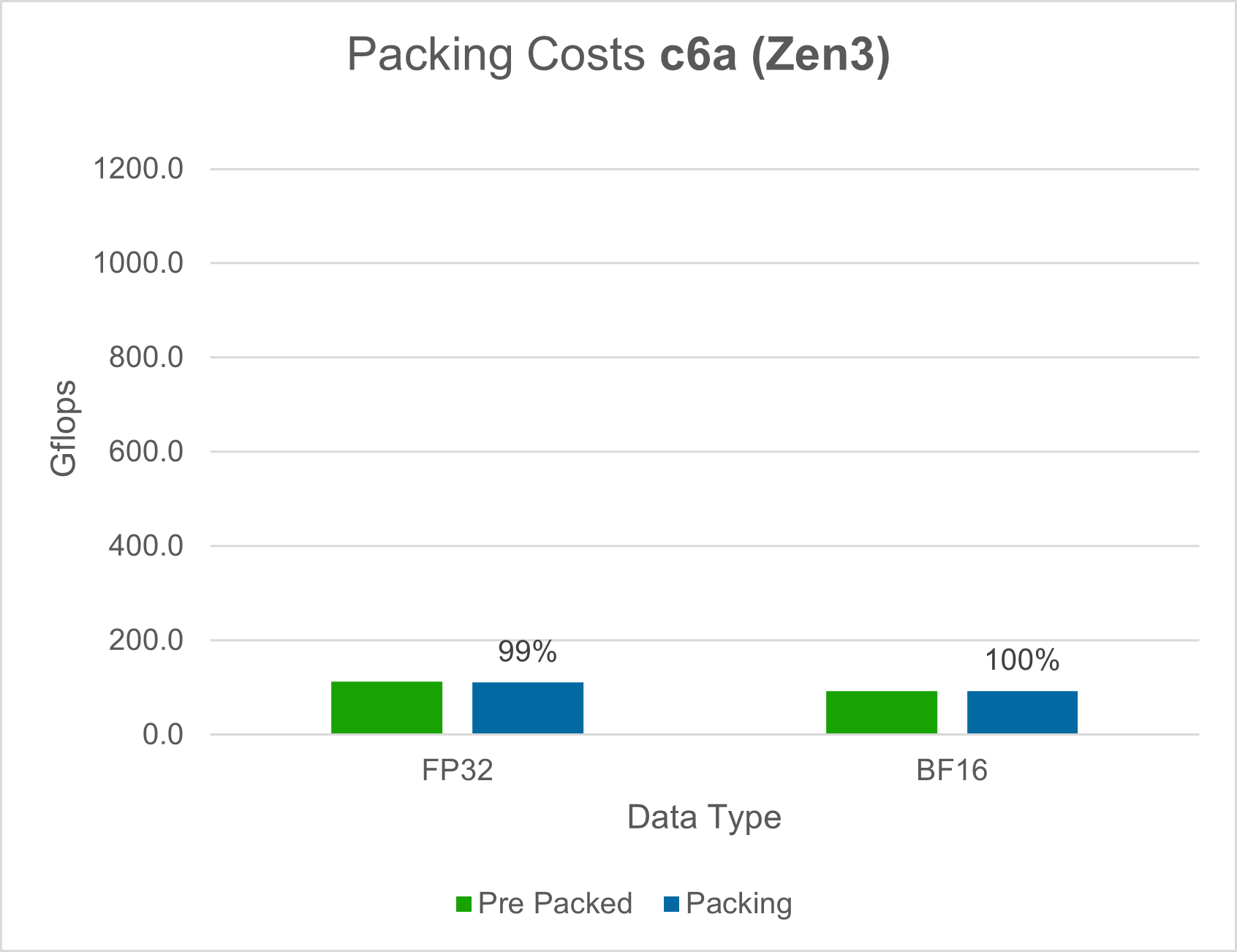}
    \end{minipage}
    \hspace{\fill} % note: no blank line here
    \begin{minipage}{0.24\textwidth}
    \includegraphics[width=1\linewidth]{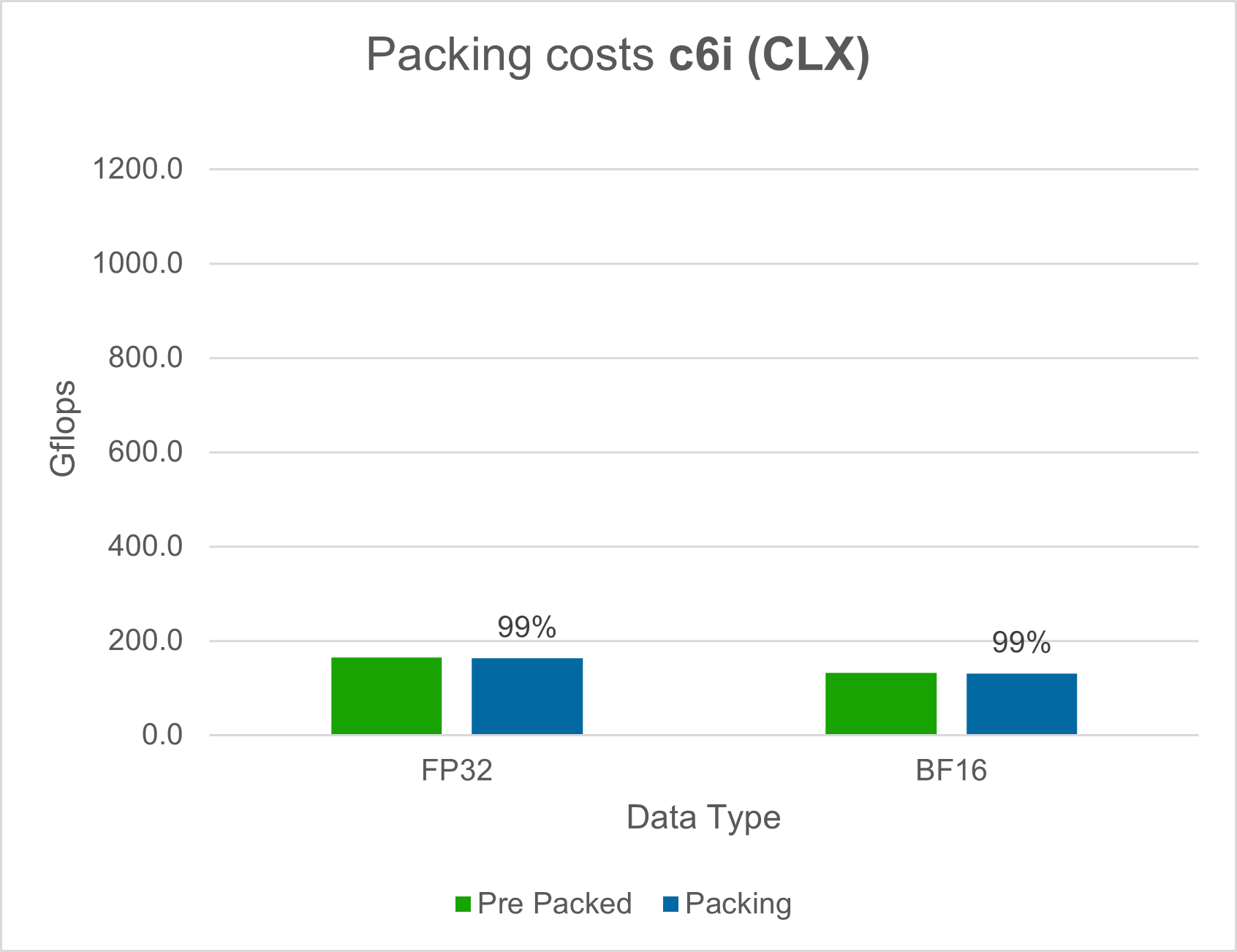}
    \end{minipage}
    \hspace{\fill} % note: no blank line here
    \begin{minipage}{0.24\textwidth}
    \includegraphics[width=1\linewidth]{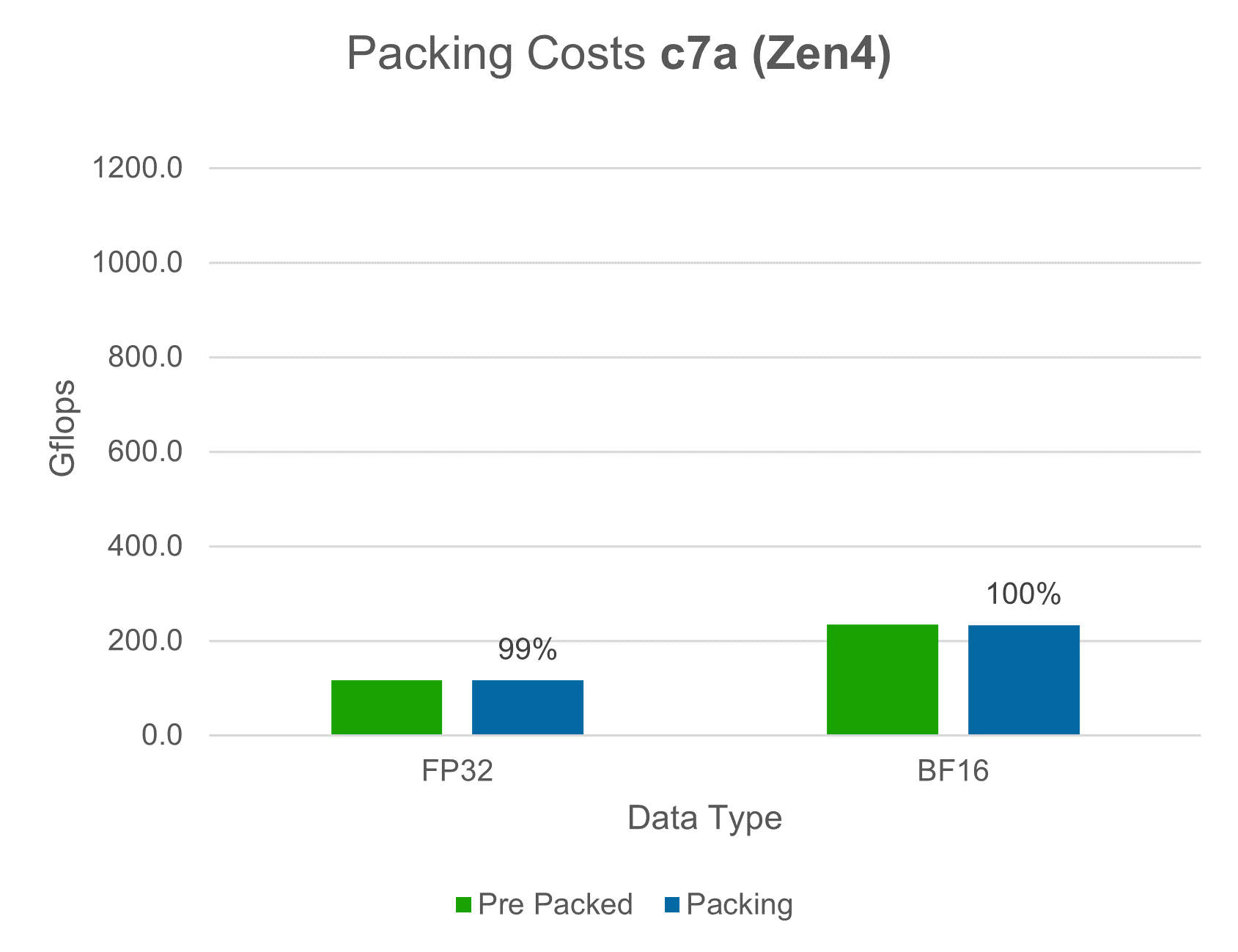}
    \end{minipage}
    \hspace{\fill} % note: no blank line here
    \begin{minipage}{0.24\textwidth}
    \includegraphics[width=1\linewidth]{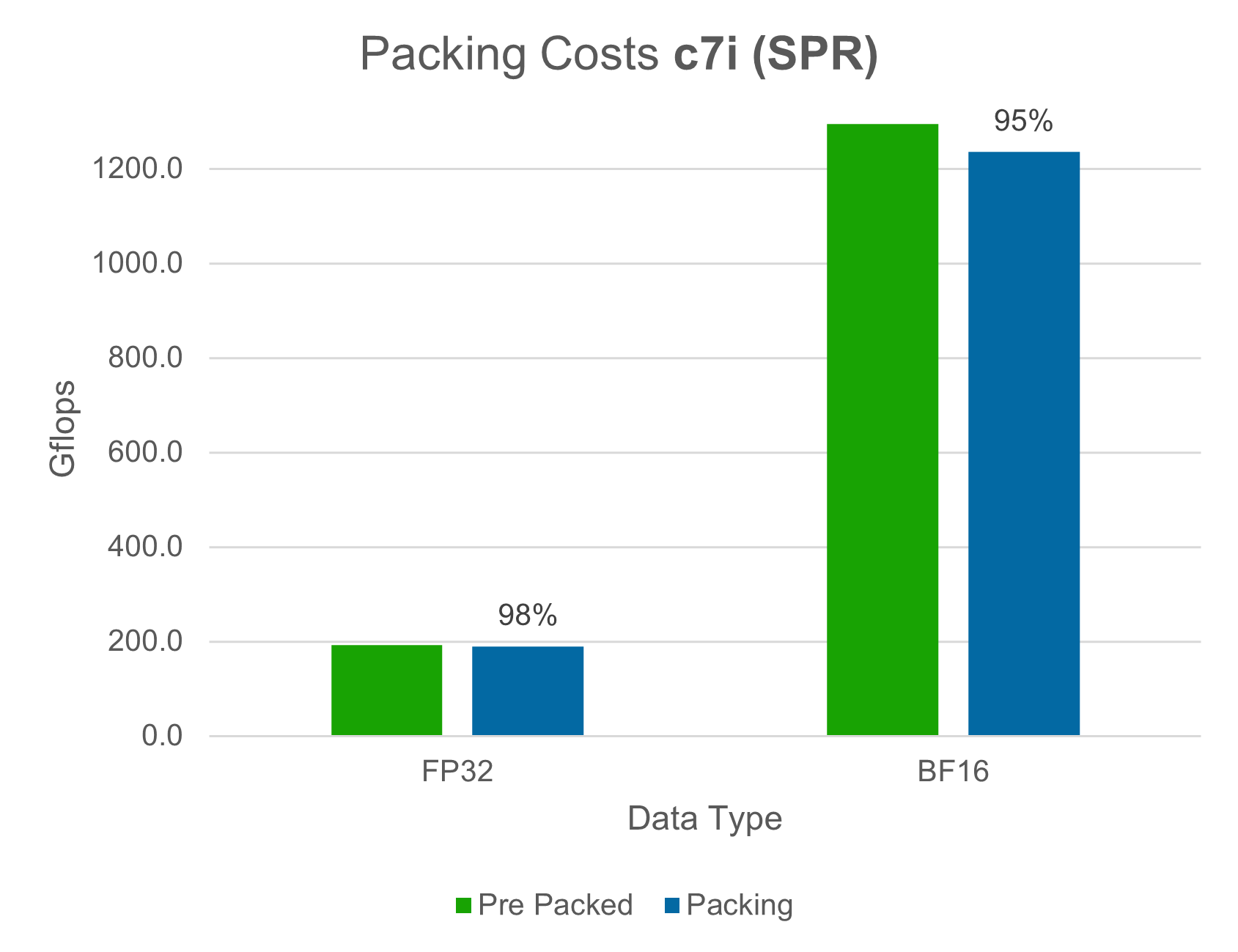}
    \end{minipage}

    \caption{Comparison of \texttt{mlir-gen} (TensorFlow-like) IR with and without compiler packing applied. This shows that the cost of packing can be easily hidden with constant packing and fusion with consumers and producers.}
    \label{fig:packing-costs}
\end{figure}

\subsection{Single-thread packing costs}

Figure \ref{fig:packing-costs} compares two types of runs:
\begin{enumerate}
    \item \textbf{Pre Packed}, where \texttt{mlir-gen} creates the model shapes in their already packed (4D) versions. This is the \texttt{TPP-MLIR} results above and are our baseline.
    \item \textbf{Packing}, where \texttt{mlir-gen} creates the model shapes in their original (2D) versions, using our compiler to find and perform tensor packing (at compile or run time).
\end{enumerate}

These results show the impact of compile and run time packing performed by the compiler, which are on average 1\% for all machines and data types. Sapphire Rapids (c7i) on BF16 shows a greater impact (7\%) due to the compute density problem identified above.

\begin{figure} [hb]
    \begin{minipage}{0.24\textwidth}
    \includegraphics[width=1\linewidth]{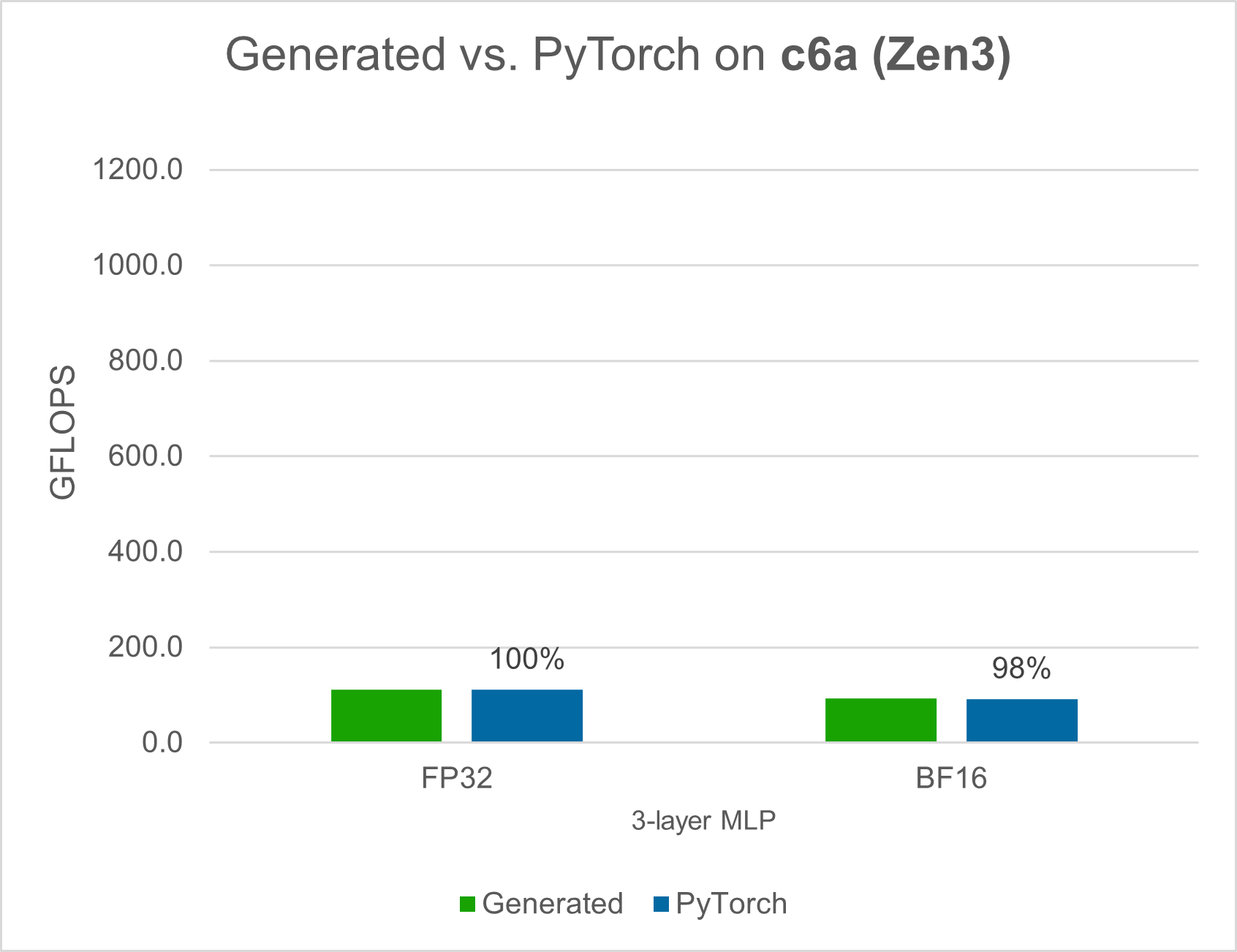}
    \end{minipage}
    \hspace{\fill} % note: no blank line here
    \begin{minipage}{0.24\textwidth}
    \includegraphics[width=1\linewidth]{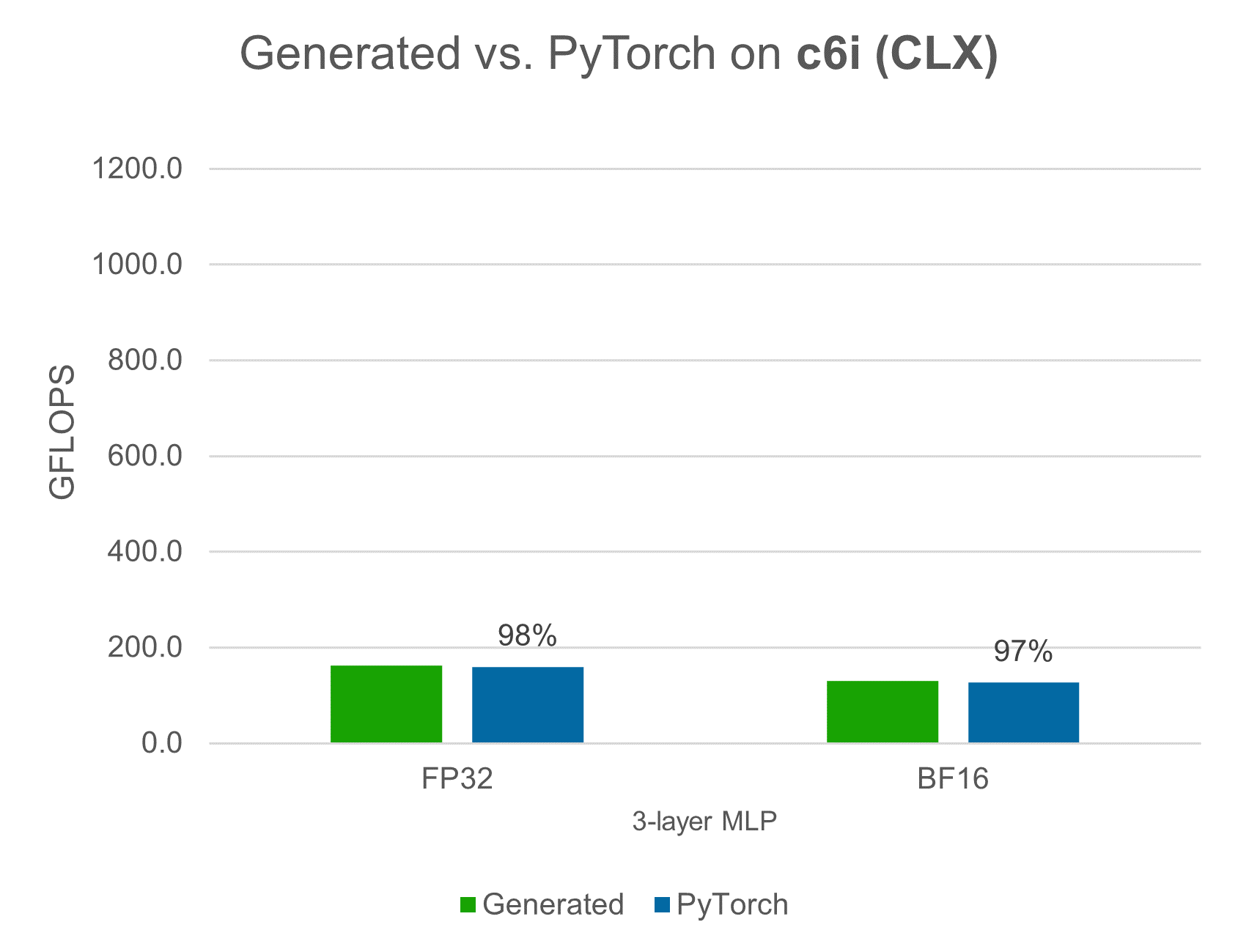}
    \end{minipage}
    \hspace{\fill} % note: no blank line here
    \begin{minipage}{0.24\textwidth}
    \includegraphics[width=1\linewidth]{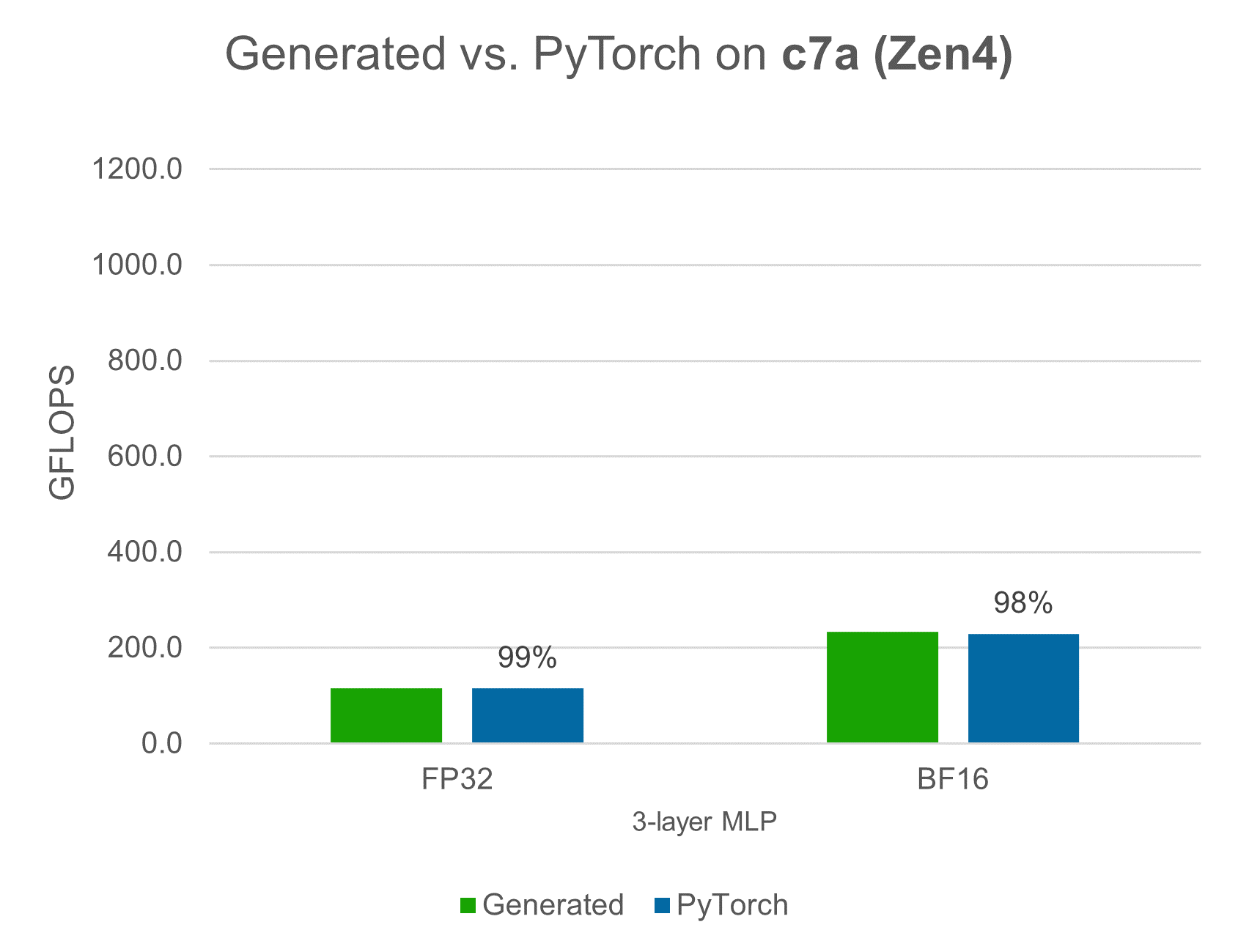}
    \end{minipage}
    \hspace{\fill} % note: no blank line here
    \begin{minipage}{0.24\textwidth}
    \includegraphics[width=1\linewidth]{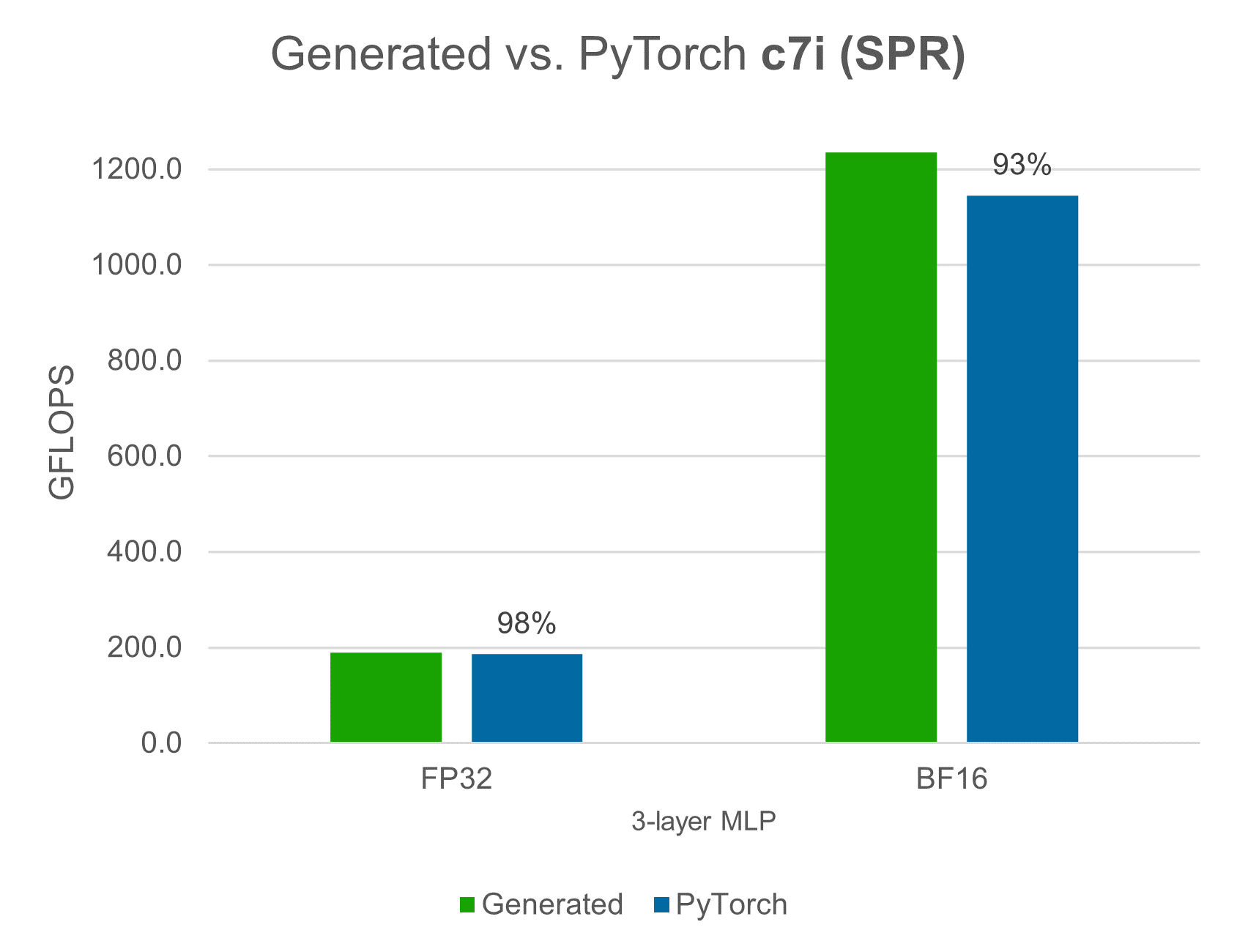}
    \end{minipage}

    \caption{Comparison of \texttt{mlir-gen} (TensorFlow-like) IR that we based our work on against PyTorch IR that we extracted through Torch Dynamo and \texttt{torch-mlir} at the end of the project.}
    \label{fig:single-thread-pytorch}
\end{figure}

\subsection{Models from new frameworks}

Figure \ref{fig:single-thread-pytorch} compares two types of runs:
\begin{enumerate}
    \item \textbf{Generated}, where \texttt{mlir-gen} creates the model shapes in their original (2D) versions, using our compiler to find and perform tensor packing. This is the \texttt{Packing} results above and are our baseline.
    \item \textbf{PyTorch}, where we extract MLIR from PyTorch on a similar kernel as the above, and pass it through our compiler. This shows how our compiler performs when encountering new IR that it hadn't seen before.
\end{enumerate}

These results show, at least for the PyTorch case, that as long as the frameworks lower the same models in similar ways, the compiler can perform the same transformations and achieve roughly the same performance (2\% on average).

The main differences between these PyTorch models and our generator were:
\begin{enumerate}
    \item TensorFlow lowers ReLU as \texttt{maxf(0, x)} while PyTorch lowers it as \texttt{max(0, x) + select(0, x)}, which needed to be recognized as an XSMM operation to lower and fuse correctly.
    \item PyTorch passes constants as arguments, so we had to force PyTorch to disable auto-grad and some other features to generate a similar kernel.
    \item PyTorch stores its weights in a transposed way, which incurs in additional memory-bound operations.
    \item PyTorch lowers the matmul and the following element wise operations writing to different buffers, so fusion does not work our of the box.
\end{enumerate}

Sapphire Rapids (c7i) on BF16 shows again a greater impact (5\%) due to the compute density problem identified above, since most of the changes are bandwidth related (optimal traversal, stray memory allocations, weight transposes).

\begin{figure} [ht]
    \begin{minipage}{0.32\textwidth}
    \includegraphics[width=1\linewidth]{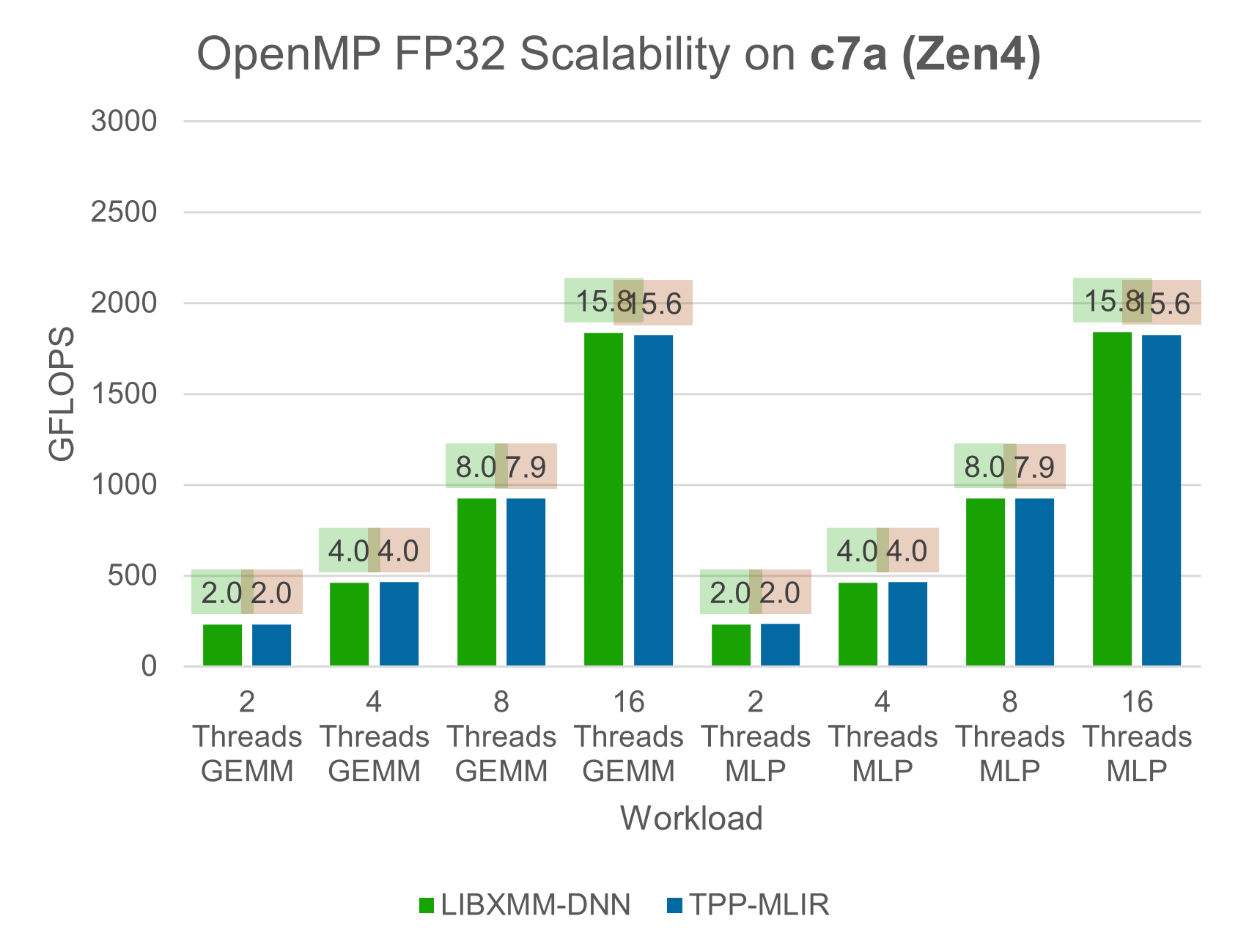}
    \end{minipage}
    \hspace{\fill} % note: no blank line here
    \begin{minipage}{0.32\textwidth}
    \includegraphics[width=1\linewidth]{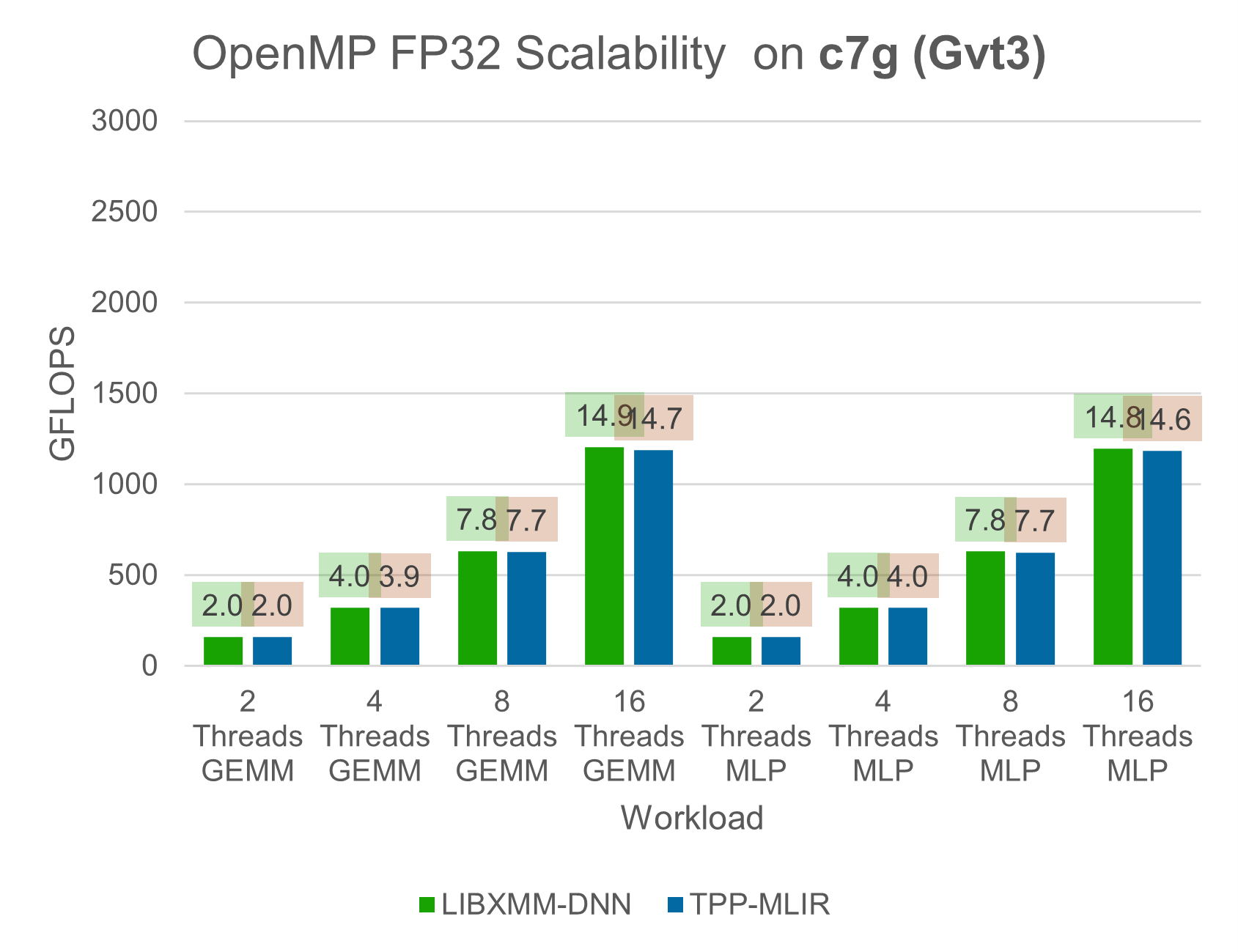}
    \end{minipage}
    \hspace{\fill} % note: no blank line here
    \begin{minipage}{0.32\textwidth}
    \includegraphics[width=1\linewidth]{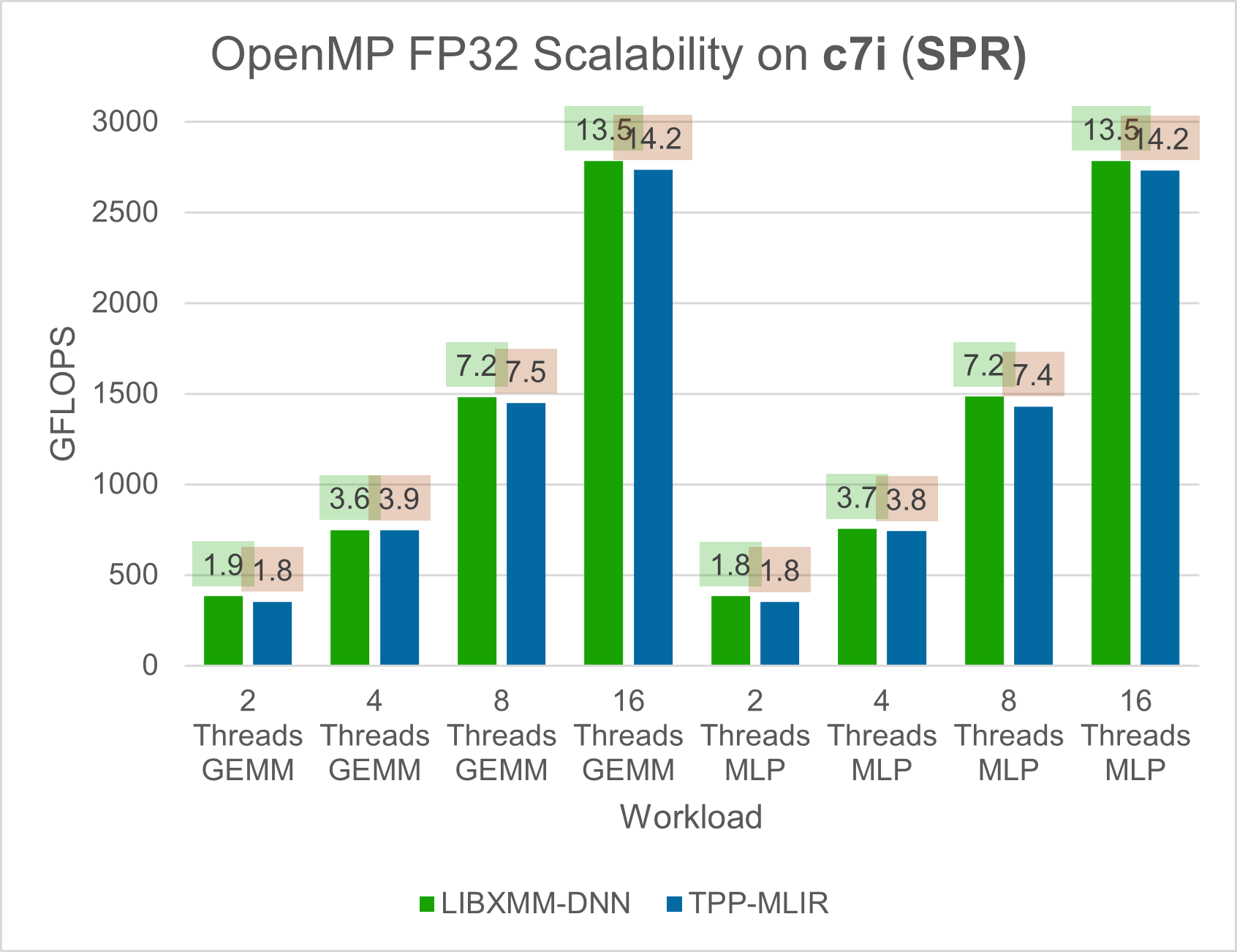}
    \end{minipage}

    \vspace*{1cm} % vertical separation

    \begin{minipage}{0.32\textwidth}
    \includegraphics[width=1\linewidth]{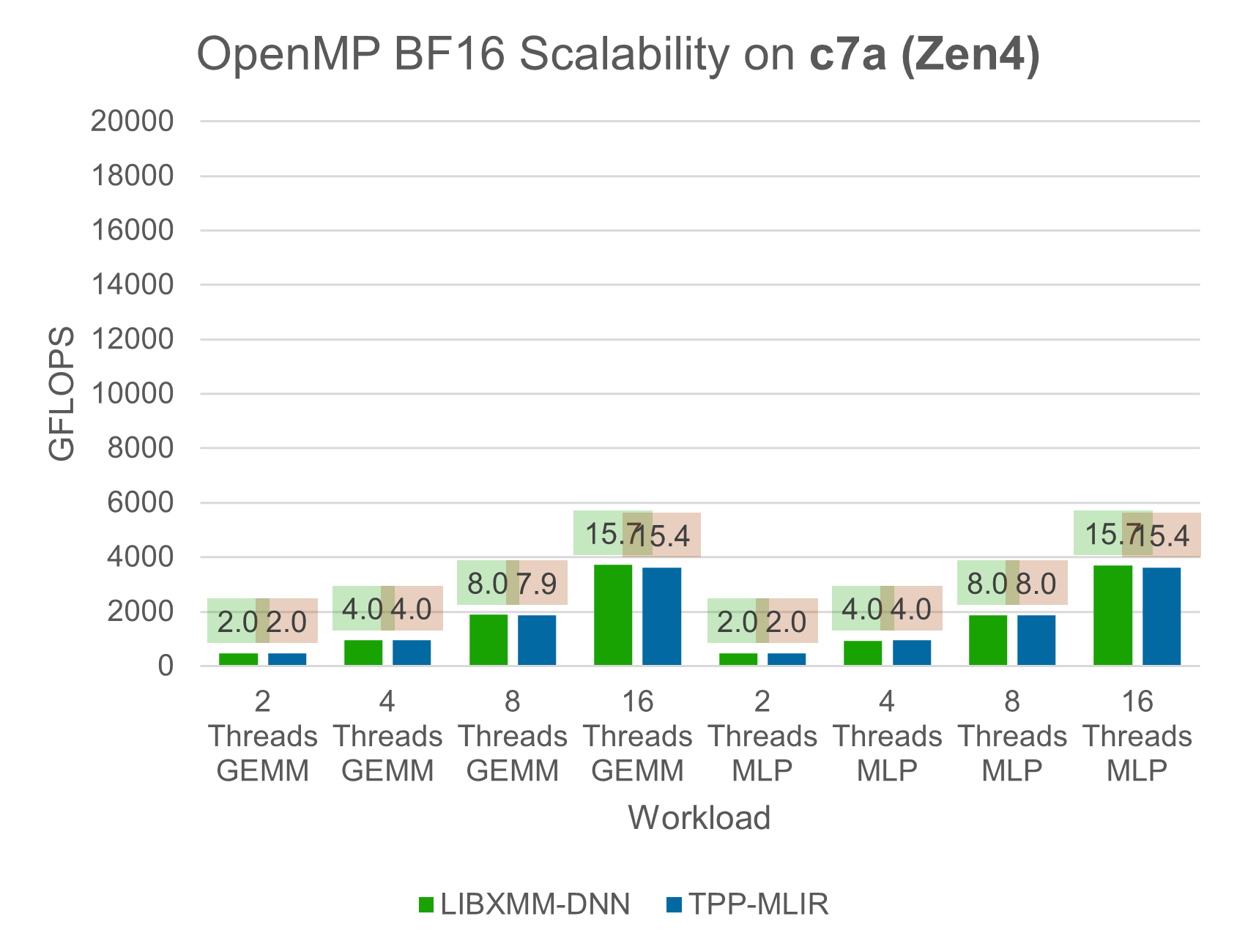}
    \end{minipage}
    \hspace{\fill} % note: no blank line here
    \begin{minipage}{0.32\textwidth}
    \includegraphics[width=1\linewidth]{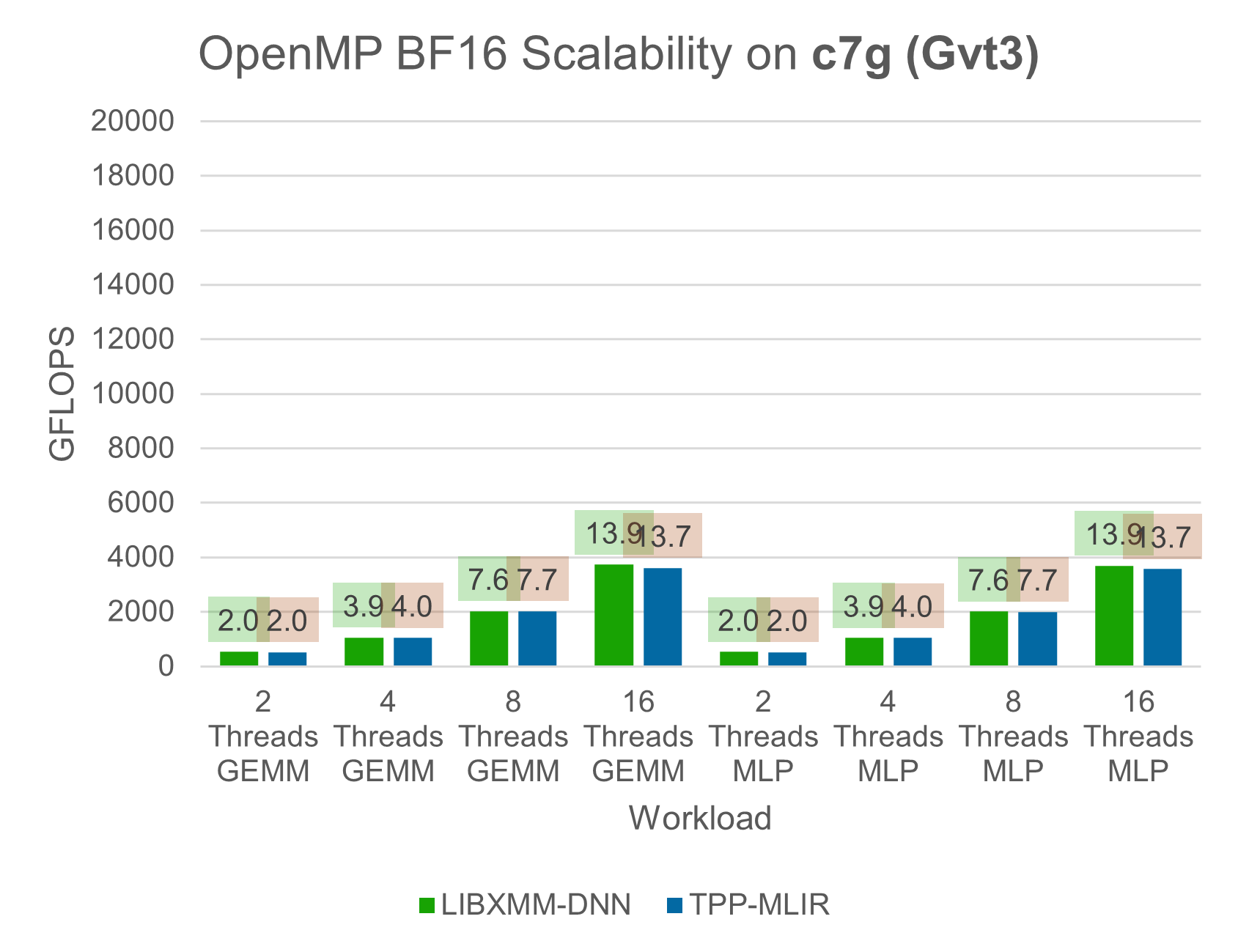}
    \end{minipage}
    \hspace{\fill} % note: no blank line here
    \begin{minipage}{0.32\textwidth}
    \includegraphics[width=1\linewidth]{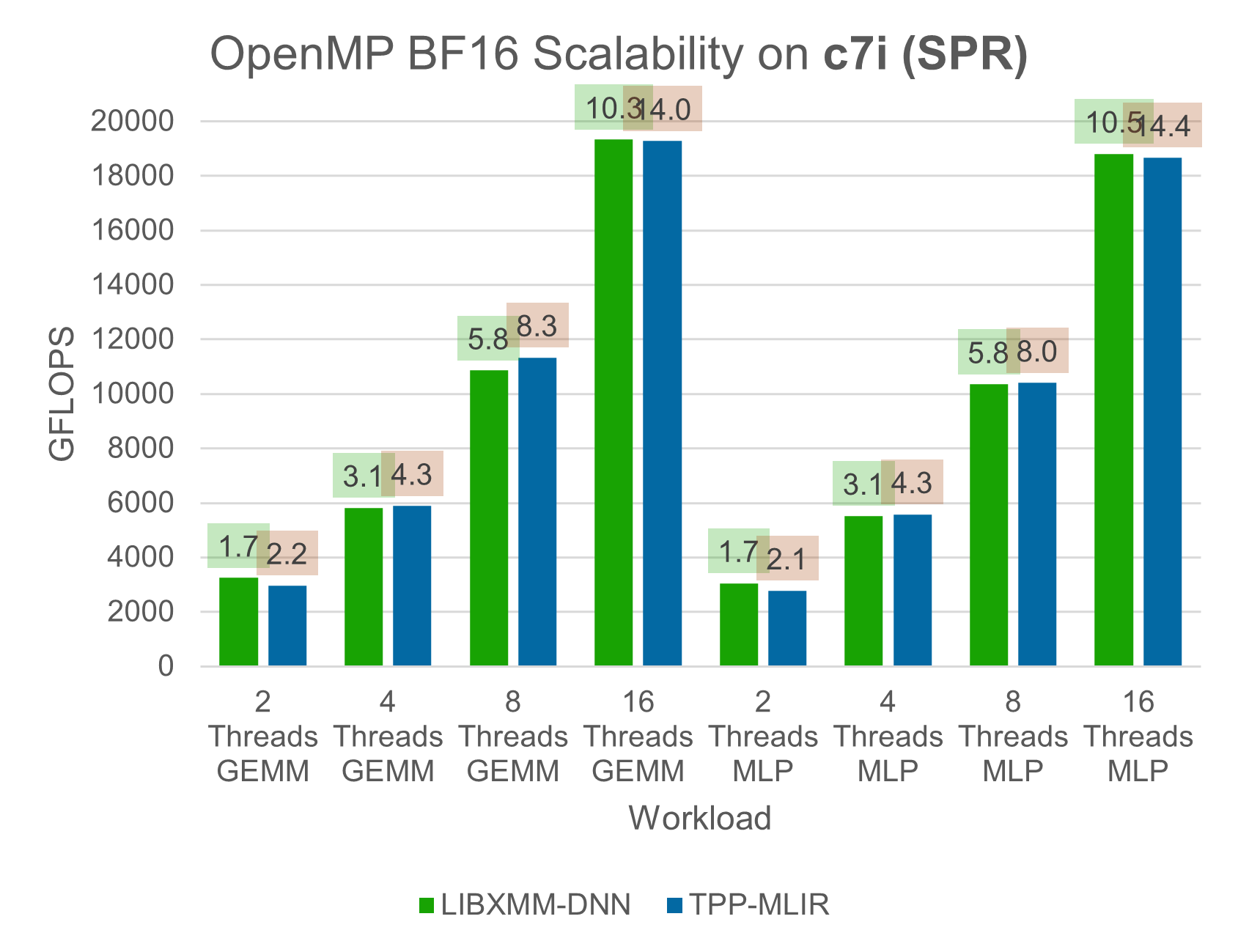}
    \end{minipage}

    \caption{Scalability of the compiler results on c7a (Zen4), c7g (Gvt2) and c7i (SPR) architectures, compared to libxsmm-dnn. The numbers on top represent speed improvement over their own single-threaded baseline. Very good scalability across the board.}
    \label{fig:multi-thread}
\end{figure}

\subsection{Parallel scalability}

Figure \ref{fig:multi-thread} compares two types of runs:
\begin{enumerate}
    \item \textbf{LIBXSMM-DNN}, our hand-written C++ code with OpenMP for multi-threaded execution. This is our baseline.
    \item \textbf{TPP-MLIR}, using our compiler on 4D shapes as above, with OpenMP for multi-threaded execution.
\end{enumerate}

We use OpenMP to test thread scalability on 2, 4, 8 and 16 threads on each hardware architecture. We use \texttt{KMP\_AFFINITY} as \texttt{granularity=fine,verbose,compact,1,0} to distribute the threads across all cores (not hyper-threads) from core 1 (since 0 is being used by the kernel).

Our results show very good scalability across the board, on both FP32 and BF16 data types. The compiler shows similar multi-threaded performance as the hand-written code, which means the same scalability on both Zen4 and Graviton 3, but increased scalability on Sapphire Rapids due to the lower single-threaded performance, while the absolute reach multi-thread performance is identical, hence the compiler delivers the same performance as the handwritten library on all platforms. 

\section{Conclusion and Future Work}

In this work, we demonstrate one can achieve comparable \textit{ninja performance} on ML models using our high-level MLIR compiler that, in turn, uses low-level tile-based building blocks without needing user input. Using a ninja-based heuristic, we have shown how to select the appropriate micro-kernels for known high-level operations.

It's important to emphasize that by upstreaming our high-level logic, we can reduce the maintenance cost for generic abstractions that are not architecture-specific, and participate in the design the the MLIR compilation infrastructure to define an industry standard, guiding design of future compilers, libraries and hardware architectures.

We have also shown the benefits of the micro-kernel approach in compilers towards separating the concerns between what a compiler can do well (high-level optimizations like tiling, fusion, and data layout) from what it is hard to achieve in practice: efficient vectorization, optimal instruction selections, pipelined register allocation. This also allows downstream experimentation on pre-production hardware extensions without having to re-write the entire stack or inject high maintenance changes into production compilers.

As future work, we plan to develop a cost model to drive the compiler heuristics, extend the programs we can optimize into more DL/HPC workloads (ex. transformers), expand the micro-kernels into other targets (GPU, accelerators), look into the interaction between micro-kernels, IR kernels and compiler vectorization on general code generations, improve the heuristics search for optimal tiling strategies and loop orders, and improve integration of downstream passes into the upstream pipeline to improve collaboration between the two worlds.

\bibliographystyle{plain}
\bibliography{references.bib}

\noindent Optimization Notice: Software and workloads used in
performance tests may have been optimized for performance only on
Intel microprocessors.  Performance tests, such as SYSmark and
MobileMark, are measured using specific computer systems,
components, software, operations and functions.  Any change to any
of those factors may cause the results to vary.  You should
consult other information and performance tests to assist you in
fully evaluating your contemplated purchases, including the
performance of that product when combined with other products.
For more information go to http://www.intel.com/performance.

\noindent Intel, Xeon, and Intel Xeon Phi are trademarks of Intel Corporation in the U.S. and/or other

\clearpage
\appendix
\section{Appendix: Reproducing results}

Our benchmarks were executed on AWS public instances:
\texttt{c6a.metal} (Zen3), \texttt{c6i.metal} (Cooper Lake), \texttt{c7i.metal-24xl} (Sapphire Rapids), \texttt{c7a.metal-48xl} (Zen4), \texttt{hpc7g.16xlarge} (Graviton 3) for the results outlined above.

Most instances are \textit{metal} instances to avoid other users interfering with our measurements.
The only one that isn't is the Graviton 3, which has scalability problems on the \textit{metal} instances but not on the HPC instances.
For this reason we reserve an entire virtualized Graviton 3 instance in order to achieve the same result and not allow other tenants on our machines.
We use a standard Amazon Linux OS on those instances.

We have fixed the state of our git repository in Github by creating a branch called \texttt{cgo-c4ml-2024}, from which these benchmarks were executed.
To reproduce the benchmark, clone our repository on that branch, enter the \texttt{scripts/benchmarks} directory and run the \texttt{build\_and\_run.sh} script, following the instructions in the README file~\footnote{https://github.com/plaidml/tpp-mlir/tree/cgo-c4ml-2024/scripts/benchmarks}.

We have used these instructions directly to run all of the numbers on this paper.

\end{document}